\documentclass[vecphys]{svmult}

\usepackage{makeidx}         % allows index generation
\usepackage{graphicx}        % standard LaTeX graphics tool
\usepackage{multicol}        % used for the two-column index
\usepackage[bottom]{footmisc}% places footnotes at page bottom
\makeindex             % used for the subject index
\begin{document}

\def\star{{\displaystyle *}}
\def\be{\begin{equation}}
\def\ee{\end{equation}}

\title*{Collective Electron Dynamics in Metallic and Semiconductor Nanostructures}
\titlerunning{Collective Electron Dynamics}
\author{G. Manfredi\inst{1}, P.-A. Hervieux\inst{1}, Y. Yin\inst{1}, \and N. Crouseilles\inst{2}}
\institute{Institut de Physique et Chimie des Mat\'eriaux de
Strasbourg,  23 Rue du Loess, BP 43, F--67034 Strasbourg, France
\texttt{giovanni.manfredi@ipcms.u-strasbg.fr} \and Institut de
Recherche en Math{\'e}matiques Avanc{\'e}es, Universit{\'e} Louis Pasteur,
Strasbourg, France \texttt{crouseil@math.u-strasbg.fr}}
\maketitle

\section{Introduction}
\label{sec:intro}

Understanding the electron dynamics and transport in metallic and
semiconductor nanostructures -- such as metallic nanoparticles,
thin films, quantum wells and quantum dots -- represents a
considerable challenge for today's condensed matter physics, both
fundamental and applied.

Experimentally, thanks to the recent development of ultrafast
spectroscopy techniques, it is now possible to monitor the
femtosecond dynamics of an electron gas confined in metallic
nanostructures such as thin films
\cite{Eesley,Brorson,Suarez,Groeneveld,Sun,Bigot,Aeschlimann,Rud03},
nanotubes \cite{Lauret}, metal clusters \cite{Schlipper,Campbell}
and nanoparticles \cite{Bigot,Aeschlimann,Voisin,Nisoli}.
Therefore, meaningful comparisons between experimental
measurements and numerical simulations based on microscopic
theories are becoming possible.

The dynamics of an electron gas confined in a metallic
nanostructure is characterised by the presence of collective
oscillations (surface plasmon) whose spectral properties depend on
several conditions of temperature, density, and coupling to the
environment. At lowest order, the linear response of the electron
gas is simply given by the plasma frequency $\omega_p =
(e^2n/m\varepsilon_0)^{1/2}$ (up to a dimensionless geometrical
factor), and does not depend on the temperature or the size of the
nano-object. The plasma frequency represents the typical
oscillation frequency for electrons immersed in a neutralizing
background of positive ions, which is supposed to be motionless
because of the large ion mass. The oscillations arise from the
fact that, when some electrons are displaced (thus creating a net
positive charge), the resulting Coulomb force tends to pull back
the electrons towards the excess positive charge. Due to their
inertia, the electrons will not simply replenish the positive
region, but travel further away thus re-creating an excess
positive charge. This effect gives rise to coherent oscillations
at the plasma frequency. Notice that, for a metallic
nanostructure, the inverse plasma frequency is typically of the
order of the femtosecond -- this coherent regime can therefore be
explored with the ultrafast spectroscopy techniques developed in
the last two decades.

The coherence of such collective motions is progressively
destroyed by Landau damping (i.e. by coupling to the internal
degrees of freedom of the electron gas) and by electron-electron
or electron-phonon collisions. The damping of the plasmon was
observed experimentally in gold nanoparticles \cite{Lamprecht} and
was studied theoretically in several works
\cite{Kreibig,Molina,Zar04}.

Although the linear response of the surface plasmon has been known
for a long time, fully nonlinear studies have only been performed
in the last decade and have revealed some interesting features.
Our own contribution to this research area has mainly focussed on
the nonlinear electron dynamics in thin metal films, where the
emergence of ballistic low-frequency oscillations has been pointed
out \cite{Manf-Herv}.

On the other hand, the same type of collective electron motion is
also observed in semiconductor nanostructures, such as quantum
wells and quantum dots. Although the spatial and temporal scales
differ by several order of magnitudes with respect to metallic
nanostructures (due the large difference in the electron density),
the relevant dimensionless parameters take similar values in both
cases \cite{Dobson}. For instance, the effective Wigner-Seitz
radius is of order unity for both metallic and semiconductor
nano-objects. Therefore, the electron dynamics can be investigated
using similar models and both types of nano-objects are expected
to share a number of similar dynamical properties.

In this review article, we will describe the collective electron
dynamics in metallic and semiconductor nanostructures using
different, but complementary, approaches. For small excitations
(linear regime), the spectral properties can be investigated via
quantum mean-field models of the TDLDA type (time-dependent local
density approximation), generalized to account for a finite
electron temperature. In order to explore the nonlinear regime
(strong excitations), we will adopt a phase-space approach that
relies on the resolution of kinetic equations in the classical
phase space (Vlasov and Wigner equations). The phase-space
approach provides a useful link between the classical and quantum
dynamics and is well suited to model effects beyond the mean field
approximation (electron-electron and electron-phonon collisions).
We will also develop a quantum hydrodynamic model, based on
velocity moments of the corresponding Wigner distribution
function: this approach should lead to considerable gains in
computing time in comparison with simulations based on
conventional methods, such as density functional theory (DFT).

The above studies all refer to the {\em charge} dynamics in a
semiconductor or metallic nanostructure, which has been
intensively studied in the last three decades. In more recent
years, there has been a surge of interest in the {\em spin}
dynamics of the carriers, mainly for possible applications to the
emerging field of quantum computing \cite{Loss}. A promising
approach to the development of a quantum computer relies on small
semiconductor devices, such as quantum dots and quantum wells
\cite{Zoller}. To implement basic qubit operations, most proposed
schemes make use of the electron spin states, so that a thorough
understanding of the spin dynamics is a necessary prerequisite.
Nevertheless, in order to manipulate the electrons themselves, one
must necessarily resort to electromagnetic fields, which in turn
excite the Coulomb mean field \cite{Gorman,Petta}. The charge and
spin dynamics are therefore closely intertwined and both must be
taken into account for a realistic modelling of
semiconductor-based qubit operations.

The ultrafast magnetization (spin) dynamics in ferromagnetic
nanostructures has also attracted considerable experimental
attention in the last decade. Pioneering experiments
\cite{demagnet} on ferromagnetic thin films revealed that the
magnetization experiences a rapid drop (on a femtosecond time
scale) when the films is irradiated with an ultrafast laser pulse,
after which it slowly regains its original value on a time scale
close to that of the electron-phonon coupling. Despite many
attempts \cite{demagnet,Zhang,Koopmans}, a clear theoretical
explanation for these effects is still lacking. Here, we will
illustrate how this problem can be addressed using some of the
techniques developed for the electron dynamics, particularly
quantum mean-field and phase-space methods, which will be
generalized to include the spin degrees of freedom.

\section{Models for the electron dynamics}
\label{sec:models} Metallic and semiconductor nano-objects operate
in very different regimes, as the electron density is several
orders of magnitudes larger for the former. Consequently, the
typical time, space, and energy scales can be very different, as
illustrated in Table \ref{tab:1}. However, if one takes into
account the effective electron mass and dielectric constant, the
relevant dimensionless parameters turn out to be rather similar
\cite{Dobson}: for instance, from Table \ref{tab:1} we see
immediately that the ratio of the screening length ($L_{\rm
screen}=v_F/\omega_p$, where $v_F$ is the Fermi velocity) to the
effective Bohr radius $a_B=4\pi\varepsilon\hbar^2/me^2$ is of
order unity. The same happens for the ratio of the plasmon energy
$\hbar\omega_p$ to the Fermi energy $E_F$, so that the normalized
Wigner-Seitz radius $r_s$ is also of order unity for both cases.
\footnote{For a quantum well, all relevant lengths far exceed the
semiconductor lattice spacing $a_{\rm latt}\simeq 5~\rm \AA$. This
makes semiconductor systems a much better approximation to jellium
(i.e., a continuum ionic density profile) than simple metals, for
which the lattice spacing is comparable to the other electronic
lengths.}

\begin{table}
\centering \caption{Typical time, space, and energy scales for
metallic and semiconductor nanostructures}
\label{tab:1}       % Give a unique label
\begin{tabular}{lll}
\hline\noalign{\smallskip}
    & Metal film & Quantum well  \\
\noalign{\smallskip}\hline\noalign{\smallskip}
$n_e$ & $10^{28} \rm m^{-3}$ & $10^{22} \rm m^{-3}$ \\
$m$ & $m_e$ & $m_\star \simeq 0.07 m_e$ \\
$\varepsilon$ & $\varepsilon_0$ & $\varepsilon \simeq 12\varepsilon_0$ \\
$L_{\rm screen}$ & $1~\rm \AA$ & $100~\rm \AA$  \\
$\omega_{p}^{-1}$ & $1 ~\rm fs$ & $1 ~\rm ps$ \\
$E_F$ & $1 ~\rm eV $ & $1 ~\rm meV $ \\
$T_F$ & $10^{4} ~\rm K$ & $10 ~\rm K$ \\
$a_B$ & $0.529~\rm \AA$ & $100~\rm \AA$ \\
$a_{\rm latt}$ & $5~\rm \AA$ & $5~\rm \AA$ \\
$r_s/a_B$ & $ 5$ & $ 3$ \\
\noalign{\smallskip}\hline
\end{tabular}
\end{table}

\begin{figure}
\centering
\includegraphics[height=6cm]{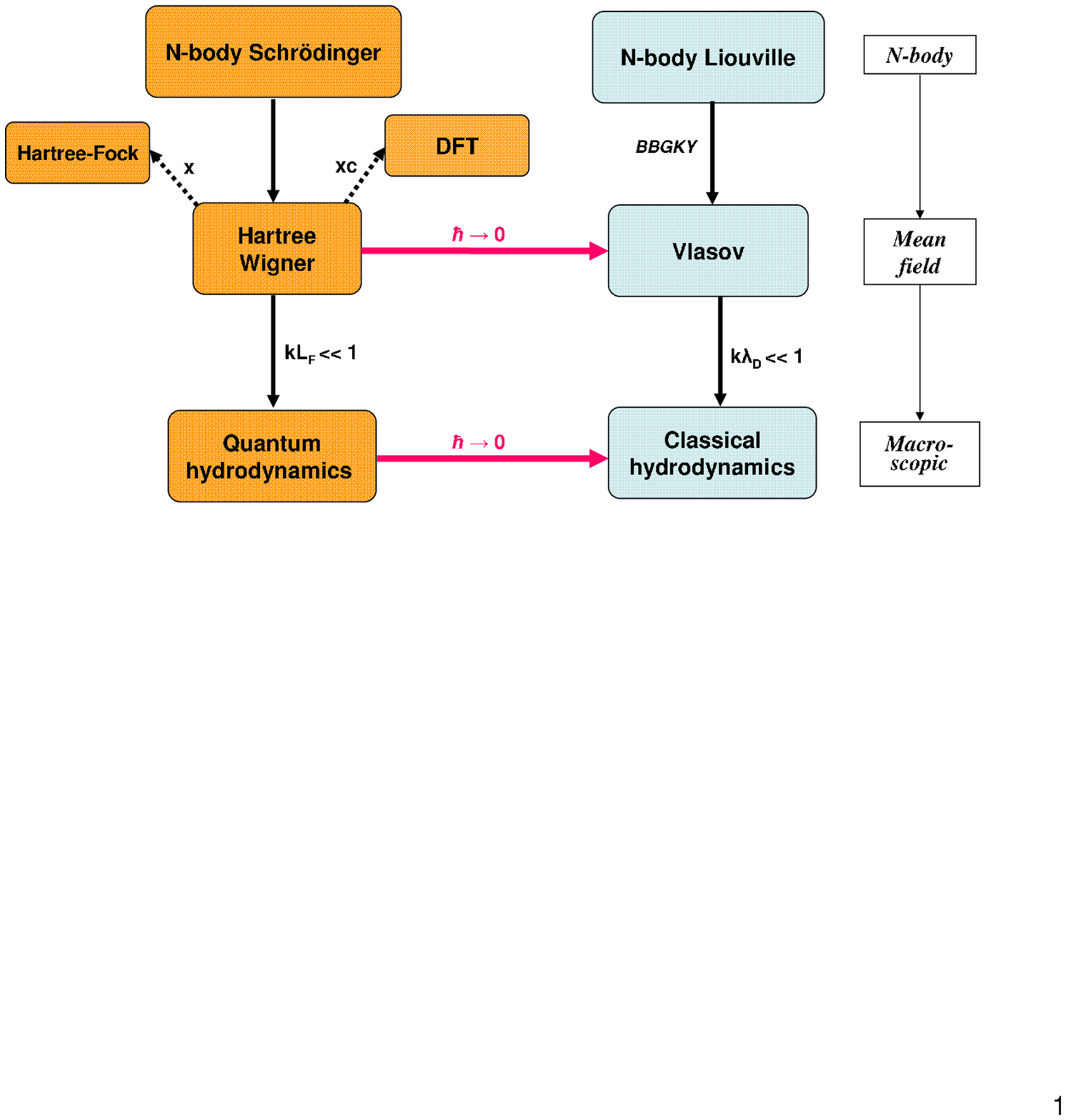}
\caption{Bird's-eye view of the models used to describe the
electron dynamics. From top to bottom: $N$-body, mean-field, and
macroscopic (hydrodynamic) theory. Left column (orange): quantum
models; right column (blue): classical models. Notation: x =
exchange; xc = exchange and correlations; $\lambda_{D}$ = Debye
length (classical screening length); $L_F$ = Thomas-Fermi
screening length; k =  typical wavevector; BBGKY = Bogoliubov,
Born, Green, Kirkwood, Yvon hierarchy.} \label{fig:schema}
\end{figure}

It is therefore not surprising that the electron dynamics of both
types of nanostructures can be described by means of similar
models. A bird's-eye view of the various relevant models is
provided in Fig. \ref{fig:schema}. The diagram represents the
various levels of modeling for the electron dynamics, both quantum
(left column, orange) and classical (right column, blue). The
highest level of description is the $N$-body model, which involves
the resolution of the $N$-particle Schr{\"o}dinger equation in the
quantum regime, or the $N$-particle Liouville equation for
classical problems (the latter is of course equivalent to Newton's
equations of motion). This is a difficult task even classically,
although molecular dynamics simulations that solve the exact
$N$-body problem can nowadays attain a considerable level of
sophistication. For Newton's equations with two-body interactions,
the numerical complexity grows at most as $N^2$, and in some cases
this can be reduced to a logarithmic dependence.
Quantum-mechanically, the $N$-body problem is virtually
unmanageable, except for very small systems, because the size of
the relevant Hilbert space grows exponentially with $N$.
Nevertheless, exact simulations of the $N$-body Schr{\"o}dinger
equation can be performed using the so-called configuration
interaction (CI) method. We have used this approach to study the
exact electron dynamics in semiconductor quantum dots containing
up to four electrons.

For larger systems, some rather drastic approximations need to be
made if we want to end up with a mathematically and numerically
tractable model. Most such reduced models are improvements on the
so-called `mean field approximation', which states that the motion
of a single electron is determined by the positions and velocities
of all other particles in the system. Such collective behavior is
possible because of the long-range nature of electromagnetic
forces. The mean field approach can be viewed as a zeroth-order
approximation to the $N$-body problem in which two-body (and
higher order) correlations between the particles have been
neglected. Classically, this procedure is known as the BBGKY
hierarchy (from the names of Bogoliubov, Born, Green, Kirkwood,
and Yvon) \cite{Balescu}.

For classical systems of charged particles (plasmas), the
mean-field dynamics is governed by the Vlasov equation, which
describes the evolution of a one-particle probability density in
the phase space. The quantum analog of the Vlasov equation is
provided by the time-dependent Hartree equations, which are
actually one-body Schr{\"o}dinger equations evolving in the mean-field
potential. In both cases, the mean field is obtained by solving
Maxwell's equations, often reduced, in the electrostatic limit, to
the sole Poisson's equation.

In this review, we concentrate on quantum mechanical models.
Several improvements have been proposed to the Hartree equations
(which were derived in 1927, just one year after Schr{\"o}dinger's
seminal paper on the wave equation), most notably Fock's
correction (1930). Indeed, the Hartree method does not respect the
principle of antisymmetry of the wavefunction, although it does
use the Pauli exclusion principle in its less stringent
formulation, forbidding the presence of two electrons in the same
quantum state. The Hartree-Fock equations respect the antisymmetry
of the wavefunctions, thus leading to an extra interaction term
between the electrons, termed the `exchange interaction'.

A particularly successful extension of the mean-field approach is
the density-functional theory (DFT), which was developed by
Hohenberg, Kohn, and Sham in the mid 1960s \cite{Kohn-Sham}.
Originally developed for the ground state at zero temperature, it
has subsequently been extended to finite temperature and
time-dependent problems. As its name suggests, DFT states that all
the properties of a many-electron systems are determined by the
electron spatial density, rather than by the wavefunctions. DFT
allows one to introduce in the mean-field formalism effects that
go beyond the strict mean-field approximation, particularly the
exchange interaction described above. Indeed, DFT can deal with
higher order correlations between the electrons, in principle
exactly if the exact density functional were known. In practice,
one has to make an educated guess for the appropriate correlation
functional, which leads to various empirical approximations.
Nevertheless, DFT has proven immensely useful for a wide range of
electronic structure calculations.

The Hartree equations can be equivalently recast in a phase-space
formalism by making use of the Wigner transformation, which was
introduced by E. Wigner in 1932 \cite{Wigner}. The resulting
Wigner function is a pseudo probability distribution, which can be
used to compute expectation values just like its classical
counterpart. Unfortunately, the Wigner function can take negative
values, which precludes the possibility of interpreting it as a
true probability density.

By taking velocity moments of the Wigner equation -- and using
some appropriate closure hypotheses -- one can derive a set of
quantum hydrodynamical (or fluid) equations that govern the
evolution of macroscopic quantities such as the particle density,
average velocity, pressure, heat flux etc. Compared to the Wigner
approach, the hydrodynamical one is obviously numerically
advantageous, as it requires the resolution of a small number of
equations in real (not phase) space. Generally speaking,
hydrodynamical methods yield accurate results over distances that
are larger than the typical electrostatic screening length, which
is the Debye length $\lambda_{D}=(k_B T_e\varepsilon/e^2 n)^{1/2}$
for classical plasmas and the Thomas-Fermi screening length
$L_F=v_F/\omega_{p}$ for degenerate electron gases (see Table
\ref{tab:1}).

In the following subsections, we shall present a brief overview of
most of the quantum models featuring in Fig. \ref{fig:schema}.

\newpage
\newpage

\subsection{Exact $N$-body simulations: the Configuration Interaction (CI) method}
\label{sec:CI} \subsubsection{Method}

In the Hartree-Fock model (HF), the many-body wave-function is
approximated by a single Slater determinant leading to a
correlation between electrons having the same spin. However,
electrons of different spin are not correlated in this
approximation. This is why the difference between the exact value
of the energy, and the HF value is called the correlation energy.
There are a number of quantum chemistry methods, which attempt to
improve the description of the many-body wave-function. The most
important one is the so-called configuration interaction method
(CI) \cite{Sherrill} which is based on the minimization of the
energy with respect to the expansion coefficients of a trial
many-body wave-function expressed as a linear combination of
Slater determinants. With respect to the models based on density
functional methods the drawback of the CI method is its
unfavorable scaling with the system size. Indeed, the dimension of
a full CI calculation grows factorially with the number of
electrons and basis functions.

From the above considerations, it is clear that CI calculations
are restricted to confined systems with very few electrons
(typically less than 10). In quantum chemistry, the "basis set"
usually refers to the set of (nonorthogonal) one-particle
functions used to build molecular orbitals. Concerning the
computational methodology for confined electron systems (atoms,
molecules, clusters, nanoparticles, quantum dots...) localized
basis sets are the traditional choice and the most common type of
basis functions is the Gaussian functions. It is worth noticing
that, from the knowledge of the exact many-body wave-function, one
can in principle: (i) compute the temporal evolution of the
system, including the dynamical correlations; (ii) obtain the true
excited states of the system.

In the following, an application of the CI method in the field of
semiconductor nanostructures and quantum dots is presented.

\subsubsection{Application}
Recent progress in semiconductors technology allows the
realization of quantum systems composed of a small number of
electrons (even a single electron!) confined in nanometer-scale
potential wells. These systems, which provide highly tunable
structures for trapping and manipulating individual electrons, are
often named artificial atoms or quantum dots and are good
candidates for the emerging technology of quantum computing. They
have certain similarities with atoms in the sense that they have a
discrete electronic structure that follows the well-known Hund's
rule of atomic physics. However, in quantum dots the electrons are
generally confined by harmonic or quasi-harmonic potentials,
whereas atoms are characterized by Coulomb confinement potentials.
The spectral properties of quantum dots are exotic with respect to
the properties of atoms in the sense that most of the oscillator
strength is concentrated almost exclusively on one dipolar
transition. This property is a direct consequence of Kohn's
theorem (KT) and does not depend on the number of electrons, the
strength of the confinement or the electron-electron interaction
\cite{Kohn}.

In a recent work \cite{Sako}, we have investigated
quasi-two-dimensional Gaussian quantum dots containing up to four
electrons within the framework of the CI method which allows in
principle an exact treatment of the many-electron system. The
Schr{\"o}dinger equation for $N$-electrons confined by a potential
$V_{ext}$ is given by
\begin{equation}
H \Psi(1,...,N) = E \Psi(1,...,N)
\end{equation}
where $(1,...,N)$ represents the space $[\vec{r}_i=(x_i,y_i,z_i)]$
and spin coordinates of the electrons and
\begin{equation}
H = \sum_{i=1}^{N}-\frac{\hbar^2}{2m}\nabla_i^2 +
\sum_{i>j}^{N}\frac{e^2}{4\pi \epsilon |\vec{r}_i-\vec{r}_j|}
+\sum_{i=1}^{N} V_{ext}(\vec{r}_i)\;.
\end{equation}
The confinement is modelled by an external one-particle
anisotropic Gaussian potential given by
\begin{equation}
V_{ext}(\vec{r}_i)=-D\exp\left[-\gamma
(x_i^2+y_i^2)\right]+\frac{1}{2}m^2 \omega_z^2 z_i^2
\;.\label{gauss}
\end{equation}
It is worth noticing that for sufficiently large values of
$\omega_z$ the electrons of the system are strongly compressed
along the $z$ direction. Therefore, in this situation, the system
can be regarded as a quantum system confined by a two-dimensional
Gaussian-type potential, i.e., as a quasi-two-dimensional Gaussian
quantum dot. Since a Gaussian potential can be approximated close
to its minimum by an harmonic potential, the potential of Eq.
(\ref{gauss}) is suitable for the modelling of anharmonic quantum
dots. The anharmonicity of the confinement can be characterized by
the depth of the Gaussian potential $D$ and by the quantity
$\omega=\sqrt{2D\gamma}/m$. Thus, when $D$ is much larger than
$\hbar \omega$ the Gaussian potential has many bound states and
the potential curve follows closely the harmonic oscillator
potential leading to a small anharmonicity of the system. On the
other hand, when $D$ is slightly larger than $\hbar \omega$ the
Gaussian potential has only few bound states and, therefore,
deviates strongly from the harmonic potential leading to a large
anharmonicity. Also, a large (small) value of $\omega$ corresponds
to a strong (weak) confinement with respect to the
electron-electron interaction.

The wave-function is approximated by a linear combination of
cartesian anisotropic Gaussian-Type Orbitals (c-aniGTO)
\cite{Diercksen}. A c-aniGTO centered at $(b_x,b_y,b_z)$ is
defined as
\begin{equation}
\chi^{\vec{a},\vec{\zeta}}(\vec{r},\vec{b})=x_{b_x}^{a_x}
y_{b_y}^{a_y} z_{b_z}^{a_z} \exp(-\zeta_x x_{b_x}^{2}-\zeta_y
y_{b_y}^{2}-\zeta_z z_{b_z}^{2})\,
\end{equation}
where $x_{b_x}=(x-b_x)$ etc...Following the quantum chemical
convention the orbitals are classified as $s$-type, $p$-type, for
$l=a_x+a_y+a_z=0,1,...,$ respectively (this sum controls the value
of the orbital angular momentum). The $(b_x ,b_y ,b_z)$ parameters
have been chosen to coincide with the center of the confining
potential. This type of basis sets was found to be the most
suitable one for expanding the eigenfunctions of an electron in an
anisotropic harmonic oscillator potential. The calculations have
been performed using the OpenMol Program \footnote{see
http://www.csc.fi/gopenmol}.

Energy spectra and oscillator strengths have been calculated for
different strength of confinement $\omega$ and potential depth
$D$. The effect of the electron-electron interaction on the
distribution of oscillator strengths and the breakdown of the KT
has been examined by focusing on the results with the same value
of $D/\hbar \omega$ i.e. with the same anharmocity.

A substantial red-shift has been observed for the dipole
transitions corresponding to the excitation into the
center-of-mass mode. The oscillator strengths, which are
concentrated exclusively in the center-of-mass excitation in the
harmonic limit, are distributed among the near-lying transitions
as a result of the breakdown of the Kohn's theorem. The
distribution of the oscillator strengths is limited to the
transitions located in the lower-energy region when $\omega$ is
large (i.e. for strongly confined electrons) but it extends
towards the higher-energy region when $\omega$ becomes small (i.e.
for weakly confined electrons).

The analysis of the CI wave functions shows that all states can be
classified according to the polyad quantum number $v_p$
\cite{Sako}. The distribution of the oscillator strengths for
large $\omega$ occurs among transitions involving excited states
with the same value of $v_p$ as the center-of-mass excited state,
$v_{p,cm}$, while it occurs among transitions involving the
excited states with $v_p=v_{p,cm}$ and $v_p=v_{p,cm+2}$ for small
$\omega$.

\subsection{Time-Dependent Density Functional Theory (TDDFT)
and the Local-Density Approximation (LDA)} \label{sec:DFT}
Time dependent density functional theory (TDDFT) extends the basic
ideas of static density functional theory (DFT) to the more
general situation of systems under the influence of time dependent
external fields. This dynamical approach relies on the electron
density $n(\vec{r},t)$ rather than on the many-body wave function
$\Psi(\vec{r}_1, \vec{r}_2,...,\vec{r}_N, t)$ of the system. In
fact, the central theorem of the TDDFT is the Runge-Gross theorem
\cite{Runge-Gross,Casida,Gross-Dobson} which tells us that all
observables are uniquely determined by the density.

From the computational point of view, with respect to the
resolution of the time dependent Schr{\"o}dinger equation (TDSE) of an
$N$-electron system, the complexity is strongly reduced when using
TDDFT. Indeed, the wave function depends on $3N+1$ variables
$(\vec{r}_1, \vec{r}_2,...,\vec{r}_N,t)$ while the density depends
only on 4 variables $(\vec{r},t)$. This is one of the reasons why
this method has become so popular. A practical scheme for
computing $n(\vec{r},t)$ is provided by the Kohn-Sham (KS)
formulation of the TDDFT \cite{Kohn-Sham}. In the latter,
noninteracting electrons are moving in an effective local
potential constructed in such a way that the KS density is the
same as the one of the interacting electron system. The advantage
of this formulation lies in its computational simplicity compared
to other quantum-chemical methods such as time-dependent
Hartree-Fock or configuration interaction. The KS equations read
as
\begin{equation}
i\hbar \frac{\partial}{\partial t} \phi_k (\vec{r},t)=\left(
-\frac{\hbar^2}{2m} \nabla^2 + V_{eff} (\vec{r},t)\right)\phi_k
(\vec{r},t) \label{tdks}
\end{equation}
with the KS density
\begin{equation}
n(\vec{r},t)=\sum_{k=1}^{\infty} f_k \left| \phi_k (\vec{r},t)
\right|^2
\end{equation}
where $f_k$ denotes the occupation numbers of the ground state,
and
\begin{equation}
V_{eff} (\vec{r},t)=V_{ext} (\vec{r},t) + V_{H} (\vec{r},t) +
V_{xc} (\vec{r},t) \;. \label{veff}
\end{equation}
In the above expression the first term is the external potential
(ionic potential, laser field...), the second is the Hartree
potential, which is a solution of the Poisson's equation, and the
last term is the exchange-correlation potential.

The most popular choice for $V_{xc}$ is the so-called adiabatic
local density approximation (ALDA) given by
\begin{equation}
V_{xc} (\vec{r},t) = \frac{d}{dn} \left[ n \epsilon_{xc}
(n)\right]_{n=n(\vec{r},t)}\;,
\end{equation}
where $\epsilon_{xc} (n)$ is the exchange-correlation energy
density for an homogeneous electron gas of density $n$. In this
approach, the same functional used to calculate the properties of
the ground state is employed in the dynamical simulations.

The validity of the local approximation has been discussed in many
papers and textbooks \cite{Lundqvist-March}. This approximation
works remarkably well for inhomogeneous electron systems. In
contrast, the validity of the adiabatic approximation has been
less thoroughly analyzed. Generally speaking, this approach is
expected to hold for finite systems and for processes that evolve
very slowly in time. The situation in bulk solids is more
controversial since significant deficiencies in the description of
absorption spectra have been noticed \cite{Onida}. It was shown by
Dobson \cite{Dobson94} that ALDA fulfills the Kohn theorem when
applied to a system of interacting electrons confined in an
external parabolic potential. This theorem guarantees the
existence of a collective state at the same frequency as the
harmonic potential. It corresponds to a rigid oscillation of the
many-body wavefunction around the center of the external
potential.

Only a few attempts have been made to go beyond ALDA. To date, the
most important ones are the work of Gross and Kohn
\cite{Gross-Kohn} and that of Vignale and Kohn
\cite{Vignale-Kohn}, the latter being the most promising in
particular for studying electron relaxation phenomena
\cite{Dagosta-Vignale}. Contrarily to ALDA, the approach of Gross
and Kohn, which uses a frequency-dependent parametrization of the
exchange-correlation kernel (see below), does not fulfill the Kohn
theorem \cite{Kohn,Dobson94}. This problem was further
investigated by Vignale and Kohn \cite{Vignale-Kohn}, who proposed
a new theory based on the so-called current density functional
theory (CDFT). This model is described in detail in
\cite{Vignale-Kohn97}. CDFT was originally derived by Vignale and
Rasolt \cite{Vignale-Rasolt} to describe, within the framework of
DFT, situations where strong magnetic fields and orbital currents
cannot be ignored.

Few works have been devoted to the study of the nonlinear electron
dynamics in finite metallic systems exposed to strong external
fields. Indeed, the resolution of the time-dependent Kohn-Sham
equations (\ref{tdks}) is a very difficult task particularly for
3D systems. Some pioneering work on free simple-metal clusters was
performed by E. Suraud in Toulouse and P.-G. Reinhard in Erlangen
\cite{Calvayrac}. More recently, Gervais et al. \cite{Gervais}
have investigated the same problem in 3D geometry using a
spherical basis expansion technique. This approach is restricted
to small metal clusters. The interaction of strong femtosecond
laser pulses with a C$_{60}$ molecule (which possesses 240
delocalized electrons and can therefore be considered as a
metallic nano-object \cite{Ruiz}) has been investigated in Ref.
\cite{Bauer} by employing a TDDFT approach. Still concerning the
fullerene molecule, Cormier et al. \cite{Cormier} studied
multiphoton absorption processes by solving numerically the
associated time-dependent Schr{\"o}dinger equation (TDSE) in the
single active electron (SAE) approximation. This approximation
consists in solving the equations (\ref{tdks}) by using, instead
of the time-dependent effective potential $V_{eff} (\vec{r},t)$
given in Eq. (\ref{veff}), the static effective potential of the
ground state together with the time-dependent electric potential
of the laser.

Let us now examine the linear regime, which has received much
wider attention in the past.

Under the condition that the external field is weak, the simplest
way to implement TDDFT is to work within the framework of the
linear response theory. This approximation was first introduced by
Zangwill and Soven \cite{Zangwill} in the context of atomic
physics for the study of photoionization in rare gases.
Subsequently, this formalism has been successfully extended to the
study of more and more complex electron systems: molecules
\cite{Levine}, simple metal clusters \cite{Eckardt}, noble metal
clusters \cite{Lerme}, thin metal films \cite{Eguiluz}, quantum
dots \cite{Lipparini-Serra}, and condensed phase systems
\cite{Onida}.

To date, in the field of nanoparticle physics, most applications
of the time-dependent Kohn-Sham formalism have been performed at
zero electron temperature. In order to interpret time-resolved
pump-probe experiments carried out on noble metal nanoparticles,
we have recently extended this approach to finite temperature. In
the following we provide a brief overview of the model with the
basic equations.

\subsubsection{Ground state}
The electron gas is assumed to be at thermal equilibrium with
temperature $T_{e}$. In the Kohn-Sham formulation of the density
functional theory at finite temperature within the grand-canonical
ensemble \cite{groundstate}, the ground-state electron density $n$
of an $N$-electron system is written, in terms of single-particle
orbitals $\phi_{i}$ and energies $\varepsilon _{i}$, as
\begin{equation}
n(\vec{r})=\sum_{k=1}^{\infty}f_{k}\
n_{k}(\vec{r})=\sum_{k=1}^{\infty}f_{k}\ |\phi _{k}(\vec{r})|^{2}
\label{rho}
\end{equation}
where $f_{k}=\left[ 1+\exp \left\{ (\varepsilon _{k}-\mu
)/k_{B}T_{e}\right\} \right] ^{-1}$ are the Fermi occupation
numbers and $\mu$ is the chemical potential. These orbitals and
energies obey the Schr\"{o}dinger equation
\begin{equation}
\left[ -\frac{\hbar^2}{2m}\nabla ^{2}+V_{eff}(\vec{r})\right] \phi _{i}(%
\vec{r})=\varepsilon _{i}\phi _{i}(\vec{r})\;,  \label{ks}
\end{equation}
where $V_{eff}(\vec{r})$ is an effective single-particle potential
given by
\begin{equation}
V_{eff}(\vec{r})=V_{ext}(\vec{r})+V_{H}(\vec{r}
)+V_{xc}(\vec{r})\;,  \label{eks}
\end{equation}
where $V_{ext}(\vec{r})$ is an external potential (e.g. due to the
ionic background), $V_{H}(\vec{r})$ is the Hartree potential
solution of Poisson's equation, and $V_{xc}(\vec{r})$ is the
exchange-correlation potential defined by
\begin{equation}
V_{xc} (\vec{r}) = \frac{d}{dn} \left[ n \omega_{xc}
(n)\right]_{n=n(\vec{r})}\;, \label{easxc1}
\end{equation}
where $\Omega _{xc}(n)\equiv\int
n(\vec{r})\;\omega_{xc}(n(\vec{r})) \; d\vec{r}$ is the
exchange-correlation thermodynamic potential \cite{Landau}. The
temperature appears in the self-consistent procedure only through
the occupation numbers and the exchange-correlation thermodynamic
potential.

For low temperature (i.e. $T_{e}\ll T_{F}[n(\vec{r})]$ where
$T_{F}[n(\vec{r})]=\frac{\hbar^2}{2mk_{B}}\left( 3\pi
^{2}n(\vec{r})\right)^{2/3}$ is the local Fermi temperature),
$\omega_{xc}(n)$ may be safely replaced by its value at $T_{e}=0$,
i.e. by $\epsilon_{xc}(n)$. The chemical potential is determined
self-consistently by requiring the conservation of the total
number of electrons from Eq. (\ref{rho})
\cite{Hervieux03,Hervieux04}.

%%%%%%%%%%%%%%%%%%%%%%%%%%%%%%%%%%%%%%%%%%%%%%%%%%%%%%%%%%%%%%%%%%%%

\subsubsection{Excited States}
In the usual first-order TDLDA at $T_{e}=0$ in the frequency
domain, the induced electron density $\delta n (\vec{r};\omega )$
is related to $\delta V_{ ext}(\vec{r}^{\prime };\omega)$, the
Fourier transform (with respect to time) of the external
time-dependent potential (generated, for instance, by the electric
field of a laser beam), via the relation
\cite{Zangwill,Eckardt,Petersilka}
\begin{equation}
\delta n (\vec{r};\omega )=\int \chi (\vec{r},\vec{r}^{\prime
};\omega )\ \delta V_{ext}(\vec{r}^{\prime };\omega )\
d\vec{r}^{\prime } \label{delta-rho}
\end{equation}
where $\chi (\vec{r},\vec{r}^{\prime };\omega )$ is the retarded
density correlation function or the dynamic response function. It
is possible to rewrite the induced density as
\begin{equation}
\delta n (\vec{r};\omega )=\int \chi ^{0}(\vec{r},\vec{r}^{\prime
};\omega )\ \delta V_{eff}(\vec{r}^{\prime };\omega )\
d\vec{r}^{\prime } \label{delta-rho-2}
\end{equation}
with
\begin{eqnarray}
\delta V_{eff}(\vec{r};\omega ) &=& \delta V_{ext}(\vec{r};\omega
)+\frac{e^2}{4\pi\epsilon_0}\int
\frac{\delta n (\vec{r}^{\prime };\omega )}{\left| \vec{r}-\vec{r}%
^{\prime }\right| }\ d\vec{r}^{\prime }  \nonumber \\
&+&\int f_{xc}(\vec{r},\vec{r}^{\prime };\omega )\ \delta n
(\vec{r}^{\prime };\omega )\ d\vec{r}^{\prime } \label{vtot-2}
\end{eqnarray}
where the function $f_{xc}(\vec{r},\vec{r}^{\prime };\omega )$ is
the Fourier transform of the time-dependent kernel defined by
$f_{xc}(\vec{r},t;\vec{r}^{\prime },t^{\prime })\equiv \delta
V_{xc }(\vec{r},t)/\delta n (\vec{r}^{\prime },t^{\prime })$ and
$\chi ^{0}(\vec{r},\vec{r}^{\prime };\omega )\ $is the
non-interacting retarded density correlation function. From Eqs.
(\ref{delta-rho})--(\ref{vtot-2}) we see that $\chi ^{0}$ and
$\chi $ are related by an integral equation (Dyson-type equation)
\begin{eqnarray}
\chi (\vec{r},\vec{r}^{\prime };\omega ) &=&\chi ^{0}(\vec{r},\vec{r}%
^{\prime };\omega )+\int \int \chi ^{0}(\vec{r},\vec{r}^{\prime
\prime
};\omega )  \nonumber \\
&\times &\ K(\vec{r}^{\prime \prime },\vec{r}^{\prime \prime
\prime };\omega
)\ \chi (\vec{r}^{\prime \prime \prime },\vec{r}^{\prime };\omega )\ d\vec{r}%
^{\prime \prime }d\vec{r}^{\prime \prime \prime }, \label{Dyson}
\end{eqnarray}
with the residual interaction defined by
\begin{equation}
K(\vec{r},\vec{r}^{\prime };\omega )=\frac{e^2}{4\pi\epsilon_0
|\vec{r}-\vec{r}^{\prime }|}+f_{xc}(\vec{r},\vec{r}^{\prime
};\omega ). \label{residual}
\end{equation}
In the {\it adiabatic} local-density approximation (ALDA) the
exchange-correlation kernel is frequency-independent and local,
and reduces to \cite{Zangwill, Petersilka}
\begin{equation}
f_{xc} (\vec{r},\vec{r}^{\prime }) = \frac{d}{dn} \left[ V_{xc}
(n)\right]_{n=n(\vec{r})}\delta \left( \vec{r}-\vec{r}^{\prime
}\right) \label{fxc}\;.
\end{equation}
It should be mentioned that the functional, $V_{xc}$ in the above
equation is the same as the one used in the calculation of the
ground state [see Eq. (\ref{easxc1})]. For spin-saturated
electronic systems, we have
\begin{eqnarray}
\chi ^{0}(\vec{r},\vec{r}^{\prime };\omega ) &=&2\sum_{jk}\ \left[
f_{j}^{0}-f_{k}^{0}\right] \ \frac{\phi _{j}^{*}(\vec{r})\phi _{k}(%
\vec{r})\phi _{k}^{*}(\vec{r}^{\prime })\phi _{j}(\vec{r}^{\prime })}{%
\hbar \omega -(\varepsilon _{k}-\varepsilon _{j})+i\eta }  \nonumber \\
&=&\sum_{k}^{occ}\ \phi _{k}^{*}(\vec{r})\phi _{k}(\vec{r}^{\prime
})\
G_{+}(\vec{r},\vec{r}^{\prime };\varepsilon _{k}+\hbar \omega )+\   \nonumber \\
&&\sum_{k}^{occ}\phi _{k}(\vec{r})\phi _{k}^{*}(\vec{r}^{\prime
})\ G_{+}^{*}(\vec{r},\vec{r}^{\prime };\varepsilon _{k}-\hbar
\omega )  \label{qui0T0}
\end{eqnarray}
where $\phi _{k}(\vec{r})$ and $\varepsilon _{k}$ are the
one-electron Kohn-Sham wave functions and energies, respectively.
$G_{+}$ is the one-particle retarded Green's function and
$f_{k}^{0}$ are the Fermi occupation numbers at $T_{e}=0$ K (0 or
1). All the above quantities are
obtained with the procedure described in the preceding subsection with $%
f_{k}=f_{k}^{0}$ in Eq. (\ref{rho}). In order to produce
numerically tractable results, we have added a small imaginary
part to the probe frequency, so that $\omega \rightarrow \omega
+i\delta$ with $\eta=\hbar \delta$.

At finite electron temperature, the grand-canonical
non-interacting retarded density correlation function reads
\cite{Yang}
\begin{eqnarray}
\chi ^{0}(\vec{r},\vec{r}^{\prime };\omega ;T_{e}) &=&\frac{1}{Z_{G}}%
\sum_{n,N}\exp \left\{ -\frac{1}{k_{B}T_{e}}\left[ E_{n}(N)-N\mu
\right]
\right\}  \nonumber \\
&\times &\chi _{n,N}^{0}(\vec{r},\vec{r}^{\prime };\omega ;T_{e})
\end{eqnarray}
where $Z_{G}$ is the grand-canonical partition function
\begin{equation}
Z_{G}=\sum_{n,N}\exp \left\{ -\frac{1}{k_{B}T_{e}}\left[
E_{n}(N)-N\mu \right] \right\}
\end{equation}
with $E_{n}(N)$ the energy of the state $\left| nN\right\rangle $
having $N$ electrons, $\mu $ the chemical potential and
\begin{eqnarray}
\chi _{n,N}^{0}(\vec{r},\vec{r}^{\prime };\omega ;T_{e}) &=&\sum_{m}\ \frac{%
\left\langle nN\left| \hat{n}(\vec{r})\right| mN\right\rangle \
\left\langle mN\left| \hat{n}(\vec{r}^{\prime })\right|
nN\right\rangle }{\hbar \omega
-(E_{m}(N)-E_{n}(N))+i\eta }  \nonumber \\
&-&\frac{\left\langle nN\left| \hat{n}(\vec{r}^{\prime })\right|
mN\right\rangle \ \left\langle mN\left| \hat{n}(\vec{r})\right|
nN\right\rangle }{\hbar \omega +(E_{m}(N)-E_{n}(N))+i\eta}.
\label{qui0T}
\end{eqnarray}
In the above expression $\hat{n}(\vec{r})$ is the particle density
\textit{operator} defined from the wave field operators by
\begin{equation}
\hat{n}(\vec{r})=\hat{\psi}^{+}(\vec{r})\hat{\psi}(\vec{r})
\end{equation}
with $\hat{\psi}^{+}(\vec{r})=\sum_{k}\hat{a}_{k}^{+}\ \phi _{k}^{*}(\vec{%
r})$ and $\hat{\psi}(\vec{r})=\sum_{k}\hat{a}_{k}\ \phi
_{k}(\vec{r})$. By using standard field theory techniques it is
possible to show that
\begin{eqnarray}
\chi ^{0}(\vec{r},\vec{r}^{\prime };\omega ;T_{e}) &=&\sum_{k}\
f_{k}\
\phi _{k}^{*}(\vec{r})\phi _{k}(\vec{r}^{\prime })\ G_{+}(\vec{r},\vec{%
r}^{\prime };\varepsilon _{k}+\hbar \omega ;T_{e})  \nonumber \\
&+&\sum_{k}f_{k}\ \phi _{k}(\vec{r})\phi _{k}^{*}(\vec{r}^{\prime
})\ G_{+}^{*}(\vec{r},\vec{r}^{\prime };\varepsilon _{k}-\hbar
\omega ;T_{e}) \label{qui0-T}
\end{eqnarray}
where $f_{k}=\left[ 1+\exp \left\{ (\varepsilon _{k}-\mu
)/k_{B}T_{e}\right\} \right] ^{-1}$. So far, we have assumed that
the residual interaction (\ref{residual}) is temperature
independent. This assumption is consistent with the use of
$\omega_{xc}(n)=\epsilon_{xc}(n)$ in the calculation of the
ground-state properties. Therefore, as for $T_{e}=0$, the response
function is solution of the Dyson equation (\ref{Dyson}) with
$\chi^{0}$ given by Eq. (\ref{qui0-T}).

The above formalism can be employed to compute the
photo-absorption by a metallic nanoparticle of size $R$. If the
wavelength $\lambda$ of the incoming light is such that $\lambda
\gg R$ the dipolar approximation is valid. From the
frequency-dependent dipole polarizability
\begin{eqnarray}
\alpha \left( \omega ;T_{e}\right) &=&\int \delta n (\vec{r};\omega; T_e )\ \delta V_{%
ext}(\vec{r};\omega )\ d\vec{r}
\end{eqnarray}
one obtains the dipolar absorption cross-section \cite{Mahan}
\begin{equation}
\sigma \left( \omega ;T_{e}\right) =\frac{\omega} {\varepsilon_{0}
c} \mathop{\rm Im} \left[ \alpha \left( \omega ;T_{e}\right)
\right] .
\end{equation}
As for the zero-temperature case, the dipolar absorption
cross-section fulfils the well known Thomas-Reiche-Kuhn (TRK) sum
rule
\begin{equation}
\int \sigma \left( \omega ;T_{e}\right) d \omega= \frac{2\pi^2
N}{c}\;.
\end{equation}

\subsubsection{Application to femtosecond spectroscopy}
Ultrafast spectroscopy using femtosecond laser pulses is a well
suited technique to study the electronic energy relaxation
mechanisms in metallic nanoparticles (see Refs.
\cite{Bigot,Voisin} and references therein). The experiments have
been carried out with nanoparticles of noble metals containing
several thousand atoms and embedded in a transparent matrix. By
using a time resolved pump-probe configuration it is possible to
have access to the spectral and temporal dependence of the
differential transmission $\frac{\Delta T}{T} (\tau,\omega)$,
defined as the normalized difference between the probe pulse with
and without the pump pulse. This quantity contains the information
on the electron dynamics, which is measured as a function of the
pump-probe time delay $\tau$ and of the laser frequency $\omega$.

For pump-probe delays longer than a few hundred femtoseconds, the
thermalization of the electrons is achieved, thus leading to an
increase of the electron temperature of several hundred degrees.
However, the electronic distribution is not in thermal equilibrium
with the lattice, the thermal relaxation to the lattice being
achieved in a few picoseconds via electron-phonon scattering. The
energy exchange between the electrons and the lattice can be
described by the two temperature model leading to a time-dependent
electron temperature $T_e(t)$ \cite{Fujimoto}
\begin{eqnarray}
C_e \frac{\partial T_e}{\partial t}&=&-G (T_e-T_i) +P(t)  \nonumber \\
C_i \frac{\partial T_i}{\partial t}&=&G (T_e-T_i),
\end{eqnarray}
where $P(t)$ represents the laser source term, $C_i$ ($C_e$) is
the lattice (electron) heat capacity, and $G$ is the
electron-lattice coupling factor. In this simplified model, the
two temperatures are assumed to be spatially uniform and therefore
the heat propagation is neglected.

Provided that the relative changes of the dielectric function with
respect to a non-perturbed system are weak (linear regime) and
that they are only due to a modification of the electron
temperature, one may identify the spectral dependence of the
differential transmission measured for a given time delay as the
difference of the linear absorption cross-sections evaluated at
different electron temperatures. More precisely, the differential
transmission is expressed as
\begin{eqnarray}
\frac{\Delta T}{T}(\tau,\omega)
&=&\frac{T[T_{e}(\tau),\omega]-T[T_e(0),\omega
]}{T[T_{e}(0),\omega]}
=-\Delta \tilde{\alpha}(\omega)\ l\label{deltaT}\\
&=&\frac{3}{2\pi R^{2}}\left[ \sigma \left( \omega ;
T_{e}(0)\right) -\sigma \left( \omega ; T_{e}(\tau)\right) \right]
\label{deltaT}
\end{eqnarray}
where $l=2R$ is the sample thickness (here, the diameter of the
nanoparticle), $T[T_e(\tau),\omega]$ and $T[T_{e}(0),\omega]$ are
the probe transmissions in the presence and absence of the pump,
respectively, and $\Delta \tilde{\alpha}$ is the pump-induced
absorption change. Obviously $T[T_e(0),\omega]$ corresponds to an
absorption at room temperature $T_e(0) = 300$ K for the conditions
where the pump-probe experiments have been performed.
\begin{figure}
\centering
\includegraphics[height=6cm]{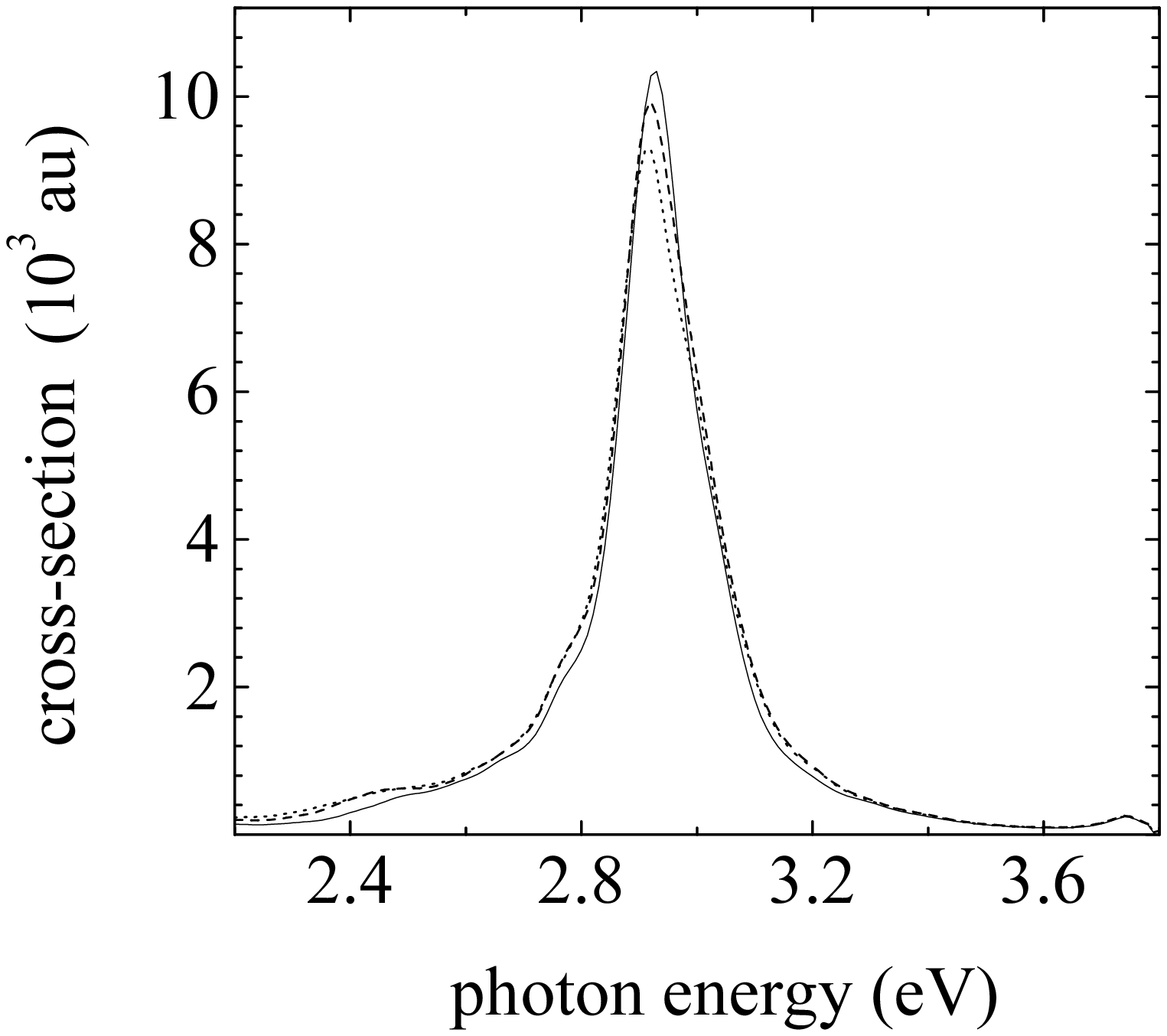}
\caption{TDLDA photoabsorption cross-section (in atomic units) of
Ag$_{2998}$ encapsulated in a transparent matrix
($\varepsilon_m=1.5$) as a function of the photon energy. Solid
line: $T_e=0$ K; dashed line: $T_e=300$ K; dotted line: $T_e=1200$
K.} \label{photoabs}
\end{figure}
We have computed the optical spectrum of a closed-shell
nanoparticle Ag$_{2998}$ embedded in a transparent matrix (alumina
$\epsilon _{m}=1.5$) for three values of the temperature. The
diameter of the nanoparticle is $4.6$ nm and the photon energy
ranges from $2.2$ eV to the interband threshold energy at $3.8$
eV, i.e. in the spectral region associated to the surface plasmon
of Ag nanoparticles. All these values correspond to typical
experimental conditions performed in our group \cite{Bigot}. The
results are presented in Fig. \ref{photoabs}. The calculated
oscillator strength is 90\%. Indeed, due to the presence of the
surface plasmon resonance, almost all the oscillator strength is
concentrated in this energy range. A clear red-shift and
broadening of the resonance as a function of the electron
temmperature is observed.

\begin{figure}
\centering
\includegraphics[height=5cm]{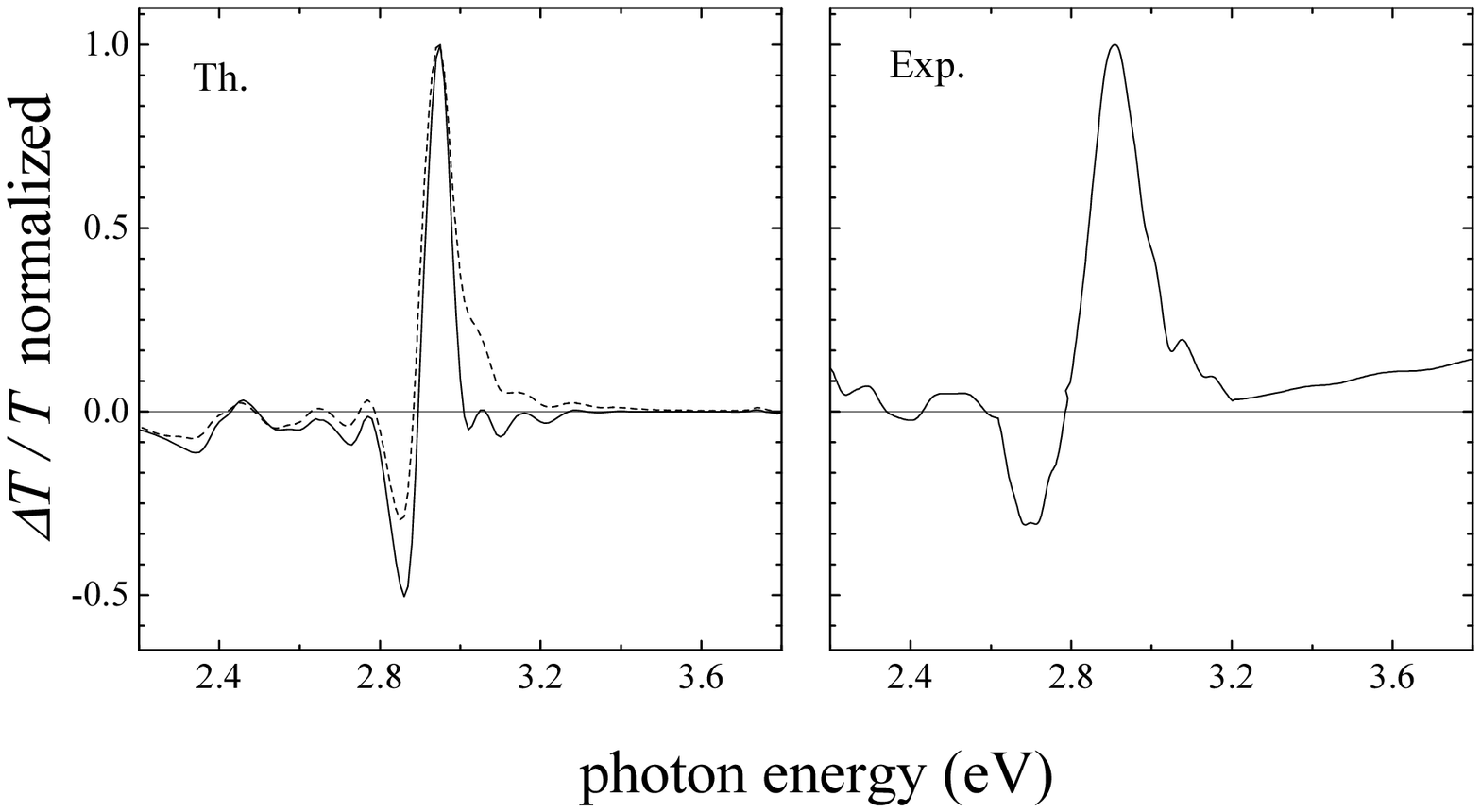}
\caption{Left panel: theoretical predictions of the normalized
differential transmission for Ag$_{2998}$ embedded in a
transparent matrix as a function of the photon energy of the
probe. Solid line: $T_e=600$ K; dotted line: $T_e=1200$ K. Right
panel: Normalized experimental spectrum of $\Delta T/T$ of silver
nanoparticles encapsulated in an alumina matrix for a pump-probe
delay of $2$ ps \cite{Bigot}.} \label{differ-transm}
\end{figure}

In the left panel of Fig. \ref{differ-transm}, the predictions of
the normalized differential transmission [Eq. (\ref{deltaT})] are
presented as a function of the photon energy of the probe. The
comparison is made for two electron temperatures $T_{e}=600$ K and
$T_{e}=1200$ K. The asymmetric shape of $\Delta T/T$ around the
resonance energy is related to a combination of a red-shift and a
broadening of the surface plasmon resonance. In the right panel of
Fig. \ref{differ-transm} the experimental spectrum of the
normalized $\Delta T/T$ obtained for a pump-probe delay of
$\tau=2$ps is depicted. The pump pulse is set at $400$ nm (second
harmonic of a titanium sapphire laser amplified at $5$ kHz) and
the probe comes from a continuum generated in a sapphire cristal
with the fundamental frequency of the amplified laser
\cite{Bigot}.

The asymmetric spectral shape of the differential transmission
spectrum in Fig. \ref{differ-transm}, which is related to the
shift and broadening of the plasmon, may have several origins. As
pointed out in Refs. \cite{Bigot,Voisin, Vallee}, the interband
transition induces a modification of the real part of the
dielectric function in this spectral region, the resonance being
far enough from the interband threshold to induce significant
changes of the corresponding imaginary part. As stressed in Refs.
\cite{Voisin, Vallee}, this is a strong indication that intraband
processes also play an important role. Indeed, as clearly seen in
Fig. \ref{photoabs}, the conduction electrons contribution leads
both to a shift and to a broadening. We can therefore conclude
that one needs to consider both the interband and intraband part
on the same footing. Whereas this effect was previously taken into
account in a phenomenological way via a shifted and broadened
Lorentzian shape, here we have derived it directly from a quantum
many-body approach based on the TDLDA at finite temperature.

\subsection{Phase space methods: from Hartree to Wigner and Vlasov}
\label{sec:phase-space} As we have seen in Sec. \ref{sec:CI}, the
most fundamental model for the quantum $N$-body problem is the
Schr\"{o}dinger equation for the $N$-particle wave function
$\Psi(\vec{r}_1,\vec{r}_2,\dots,\vec{r}_N,t)$. Unfortunately, the
full Schr\"{o}dinger equation cannot be solved exactly except for
very small systems. A drastic, but useful and to some extent
plausible, simplification can be achieved by neglecting two-body
(and higher order) correlations. This amounts to assume that the
$N$-body wave function can be factored into the product of $N$
one-body functions: \be
\Psi(\vec{r}_1,\vec{r}_2,\dots,\vec{r}_N,t) =
\psi_1(\vec{r}_1,t)~\psi_2(\vec{r}_2,t)\dots \psi_N(\vec{r}_N,t).
\label{eq:nbody} \ee
For fermions, a weak form of the exclusion principle is satisfied
if none of the wave functions on the right-hand side of Eq.
(\ref{eq:nbody}) are identical \footnote{A stronger version of the
exclusion principle requires that
$\Psi(\vec{r}_1,\vec{r}_2,\dots,\vec{r}_N,t)$ is antisymmetric,
i.e. that it changes sign when two of its arguments are
interchanged. This can be achieved by taking, instead of the
single product of $N$ wave functions as in Eq. (\ref{eq:nbody}), a
linear combinations of all products obtained by permutations of
the arguments, with weights $\pm 1$ (Slater determinant)
\cite{Ashcroft}. This is at the basis of Fock's generalization of
the Hartree model}.

When the above assumption is made, the $N$-body Schr\"{o}dinger
equation reduces to a set of one-particle equations, coupled
through Poisson's equation ({\em time-dependent Hartree model}):
\begin{eqnarray}
i\hbar\frac{\partial\psi_\alpha}{\partial\,t} &=& -
\frac{\hbar^2}{2m}~ \Delta \psi_\alpha
- e\phi\psi_\alpha~, ~~~~~ \alpha = 1 \dots N_{\rm orb} \label{eq:hartree}\\
\Delta\phi &=& \frac{e}{\varepsilon}\left(\sum_{\alpha=1}^{N_{\rm
orb}} p_\alpha |\psi_{\alpha}|^2 - n_{i}(\vec{r})\right)
\label{eq:poisson},
\end{eqnarray}
where $N_{\rm orb}\geq N$ is the number of occupied orbitals, $e$
and $m$ are the absolute electron charge and mass, and
$\varepsilon$ is the dielectric constant; $n_i(\vec{r})$ is the
ion density, which is supposed to be fixed and a continuous
function of the position coordinate. This is known as the
`jellium' hypothesis and is valid whenever the relevant length
scales are significantly larger that the ionic lattice spacing
$a_{\rm latt} \sim 5\rm\AA$. As mentioned in Sec.
\ref{sec:models}, this is the case for semiconductor
nanostructures, but not so for metals (see Table \ref{tab:1});
nevertheless, the jellium models still yields reasonably accurate
results for all but the smallest nano-objects.

The occupation probabilities $p_\alpha$ ($\sum_{\alpha=1}^{N_{\rm
orb}} p_\alpha =1$) are defined to describe a Fermi-Dirac
distribution at finite electron temperature, $p_\alpha =
[1+\exp(\beta (\epsilon_\alpha-\mu))]^{-1}$, where $\beta=1/k_B
T_e$, $\mu$ is the chemical potential, and $\epsilon_\alpha$ is
the single-particle energy level. In practice, one first needs to
obtain the ground-state equilibrium solution of Eqs.
(\ref{eq:hartree})-(\ref{eq:poisson}), which amounts to
determining the $N_{\rm orb}$ occupation probabilities and the
corresponding energy levels and wave functions. Subsequently, the
equilibrium can be perturbed to study the electron dynamics. The
numerical methods for the dynamics are quite standard, as the Eqs.
(\ref{eq:hartree}) are basically one-particle Schr{\"o}dinger
equations. We will not enter into the details of the numerical
methods in this paper: a list of relevant works on the Schr{\"o}dinger
equation can be found in Ref. \cite{Truong}.

We now show that the Hartree equations can be written in a
completely equivalent form by making use of the Wigner
transformation. The Wigner representation \cite{Wigner} is a
useful tool to express quantum mechanics in a phase space
formalism (for reviews see \cite{Moyal,Tatarskii,Hillery}). The
Wigner function is a function of the phase space variables $(x,
v)$ and time, which, in terms of the single-particle wave
functions, reads as

\be f(x,v,t) = \sum_{\alpha=1}^{N_{\rm orb}} \frac{m}{2\pi\hbar}~
p_\alpha \int_{-\infty}^{+\infty} \psi_\alpha^{\displaystyle*}
\left(x + \frac{\lambda}{2},t\right) \psi_\alpha\left(x -
\frac{\lambda}{2},t\right) e^{i m v\lambda/\hbar}~d\lambda
\label{wigfunc} \ee
(we restrict our discussion to one-dimensional cases, but all
results can easily be generalized to three dimensions). It must be
stressed that the Wigner function, although it possesses many
useful properties, is not a true probability density, as it can
take negative values. However, it can be used to compute averages
just like in classical statistical mechanics. For example, the
expectation value of a generic quantity $A(x,v)$ is defined as:
\be \langle A \rangle = \frac{\int \int f(x,v) A(x,v) dx dv}{\int
\int f(x,v) dx dv}, \label{average}\ee
and yields the correct quantum-mechanical value \footnote{For
variables whose corresponding quantum operators do not commute
(such as $\hat{x}\hat{v}$), Eq. (\ref{average}) must be
supplemented by an ordering rule, known as Weyl's rule
\cite{Hillery}.}. In addition, the Wigner function reproduces the
correct quantum-mechanical marginal distributions, such as the
spatial density: \be n(x,t) = \int_{-\infty}^{+\infty} f(x,v,t)
~dv = \sum_{\alpha=1}^{N_{\rm orb}} p_\alpha \mid \psi_\alpha
\mid^2. \ee

We also point out that, of course, not all functions of the phase
space variables are genuine Wigner functions, as they cannot
necessarily be written in the form of Eq. (\ref{wigfunc}). In
general, although it is trivial to find the Wigner function given
the wave functions that define the quantum mixture, the inverse
operation is not generally feasible. Indeed, there are no simple
rules to establish whether a given function of $x$ and $v$ is a
genuine Wigner function. For a more detailed discussion on this
issue, and some practical recipes to construct genuine Wigner
functions, see \cite{Manfredi-Feix}.

The Wigner function obeys the following evolution equation:
\[
\frac{\partial{f}}{\partial{t}} + v\frac{\partial{f}}{\partial{x}}
~+
\]
\be \frac{em}{2i\pi\hbar^2}
\int\int{d\lambda}~{dv'}e^{im(v-v')\lambda/\hbar} \label{wignereq}
\left[\phi\left(x+\frac{\lambda}{2}\right)-
\phi\left(x-\frac{\lambda}{2}\right)\right]f(x,v',t) = 0~, \ee
where $\phi(x,t)$ is the self-consistent electrostatic potential
obtained self-consistently from Poisson's equation
(\ref{eq:poisson}).

Developing the integral term in Eq. (\ref{wignereq}) up to order
$O(\hbar^2)$ we obtain \be \frac{\partial{f}}{\partial{t}} +
v\frac{\partial{f}}{\partial{x}}+ \frac{e}{m} \frac{\partial
\phi}{\partial{x}} \frac{\partial{f}}{\partial{v}} =
\frac{e\hbar^2}{24 m^3} \frac{\partial^3 \phi}{\partial{x^3}}
\frac{\partial^3{f}}{\partial{v^3}}+ O(\hbar^4). \ee
In the limit $\hbar \to 0$ one recovers the classical Vlasov
equation, well-known from plasma physics (see Fig.
\ref{fig:schema}). The Vlasov-Poisson  system has been used to
study the dynamics of electrons in metal clusters and thin metal
films \cite{Calvayrac,Manf-Herv}. It is appropriate for large
excitation energies, for which the electrons' de Broglie
wavelength is relatively small, thus reducing the importance of
quantum effects in the electron dynamics. Nevertheless, for
metallic nanostructures at room temperature, the equilibrium must
be given by a Fermi-Dirac distribution, because the Fermi
temperature is very high (see Table \ref{tab:1}). For
semiconductor nanostructures, $T_F \sim 10-50 \rm K$, so that a
Maxwell-Boltzmann equilibrium is sometimes appropriate for
moderate temperatures.

The Wigner equation must be coupled to the Poisson's equation for
the electric potential:
\be \label{eq:poisson-w} \frac{\partial^2 \phi}{\partial x^2} =-
\frac{e}{\varepsilon}\left[n_i(x) - n(x,t)\right], \ee
The resulting Wigner-Poisson (WP) system has been extensively used
in the study of quantum transport
\cite{Kluksdahl,Markowich,Drummond}. Exact analytical results can
be obtained by linearizing Eqs. (\ref{wignereq}) and
(\ref{eq:poisson-w}) around a spatially homogeneous equilibrium
given by $n_0 f_0(v)$ (Maxwell-Boltzmann or Fermi-Dirac
distribution), where $n_0=n_i=\rm const.$ is the uniform
equilibrium density. By expressing the fluctuating quantities as a
sum of plane waves $\exp(i k x -i \omega t)$ with frequency
$\omega$ and wave number $k$, the dispersion relation can be
written in the form $\varepsilon(k,\omega)=0$, where the
`dielectric constant' $\varepsilon$ reads, for the WP system,
\be \varepsilon_{\rm WP}(\omega, k) = 1 + \frac{m \omega_p^2}{n_0
k} \int \frac{f_0(v+\hbar k/2m) - f_0(v-\hbar k/2m)}{\hbar k
(\omega-k v)}~dv, \label{dispwp1}\ee
or equivalently
\begin{equation}
\label{dispwp2} \varepsilon_{\rm WP}(\omega, k) = 1 -
\frac{\omega_{p}^2}{n_{0}}\int\frac{f_{0}(v)} {(\omega - kv)^{2} -
\hbar^{2}k^{4}/4m^{2}}~dv~.
\end{equation}
This is just the Lindhard \cite{Lindhard} dispersion relation,
well known from solid-state physics. From Eq. (\ref{dispwp1}), one
can recover the Vlasov-Poisson dispersion relation by taking the
classical limit $\hbar\to 0$
\be \varepsilon_{\rm VP}(\omega, k) = 1 + \frac{\omega_p^2}{n_0 k}
\int \frac{\partial f_0/\partial v}{\omega-k v}~dv.
\label{dispvp}\ee

The equivalence of the Hartree and Wigner-Poisson methods can be
easily proven by comparing the linear results. For the Hartree
equations (\ref{eq:hartree}), we linearize around a homogeneous
equilibrium given by plane waves:
\be \psi_{\alpha} = \sqrt{n_0}~\exp\left(i \frac{m
u_{0\alpha}}{\hbar}x\right), \label{equipsi}\ee each with
occupation number $p_\alpha$ and energy
$\epsilon_\alpha=mu_{0\alpha}^2/2$. The Hartree dielectric
constant is found to be
\begin{equation}
\label{dispmulti} \varepsilon_{\rm H}(\omega, k) = 1 -
\sum_{\alpha=1}^{N_{\rm orb}} p_{\alpha}
~\frac{\omega_{p}^2}{(\omega - k u_{0\alpha})^{2} -
\hbar^{2}k^{4}/4m^{2}},
\end{equation}
which is a discrete form of the Wigner-Poisson dispersion relation
(\ref{dispwp2}).

\paragraph{Example --- Ultrafast electron dynamics in thin metal films}
Several experiments have shown \cite{Brorson,Suarez} that electron
transport in thin metal films occurs on a femtosecond time scale
and involves ballistic electrons traveling at the Fermi velocity
of the metal $v_F$. More recently, a regime of low-frequency
nonlinear oscillations (corresponding to ballistic electrons
bouncing back and forth on the film surfaces) was measured in
transient reflection experiments on thin gold films \cite{Liu}.

These findings were corroborated by accurate numerical simulations
based on the one-dimensional Vlasov-Poisson equations
\cite{Manf-Herv}. The electrons are initially prepared in a
Fermi-Dirac equilibrium at finite (but small) temperature. They
are subsequently excited by imposing a constant velocity shift
$\Delta v=0.08v_F$ to the initial distribution, which is a rather
strong excitation. This scenario is appropriate when no linear
momentum is transferred parallel to the plane of the surface
(i.e., $q_{\parallel}=0$) and is relevant to the excitation of the
film with optical pulses \cite{Anderegg}. For $q_{\parallel}=0$,
only longitudinal modes (volume plasmon with $\omega =
\omega_{p}$) can be excited.

\begin{figure}
\centering
\includegraphics[height=4.5cm, width=6cm]{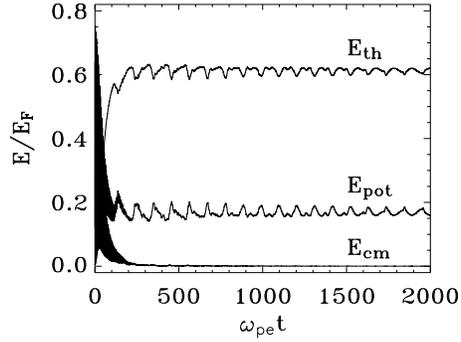} \caption{Time evolution of the
thermal, potential, and center-of-mass energies of the electron
population in a thin sodium film.} \label{fig:film}
\end{figure}

As a reference case, we studied a sodium film with initial
temperature $T_e=0.008 T_F \simeq 300$ K  and thickness $L \simeq
120~$\AA. The time evolution of the thermal $E_{\rm th}$ and
center-of-mass $E_{\rm cm}$ energies was analyzed (Fig.
\ref{fig:film}). During an initial rapidly-oscillating phase,
$E_{\rm cm}$ is almost entirely converted into thermal energy
(Landau damping). After saturation, a slowly oscillating regime
appears, with period equal to $50 \omega_{p}^{-1} \approx
5.3\rm{fs}$, where $\omega_{p}=(e^2n/m\varepsilon_0)^{1/2}$ is the
plasmon frequency. This period is close to the time of flight of
electrons traveling at the Fermi velocity and bouncing back and
forth on the film surfaces (further details are provided in our
previous work \cite{Manf-Herv}).

\begin{figure}
\centering
\includegraphics[height=7cm,width=9cm]{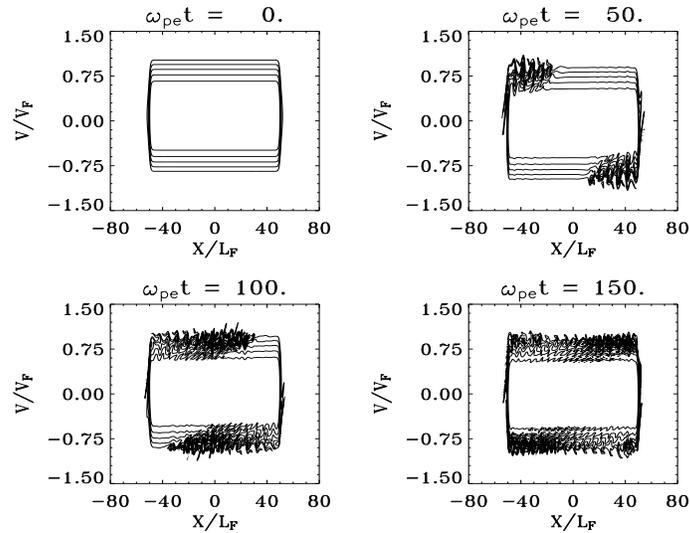} \caption{Phase space portrait of the electron
distribution. Velocity is normalized to the Fermi velocity, and
space to the Thomas-Fermi screening length $L_F=v_F/\omega_{p}$.}
\label{fig:phase-elec}
\end{figure}

The phase space portrait of the electron distribution, shown in
Fig. \ref{fig:phase-elec} clearly reveals that the perturbation
starts at the film surfaces, and then proceeds inward at the Fermi
velocity of the metal. The structure formation at the Fermi
surface, which has spread over the entire film for $\omega_{p}t >
150$, is responsible for the increase of the thermal energy (and
thus the electron temperature) observed in Fig. \ref{fig:film}. As
no coupling to an external environment (e.g., phonons) is present,
this excess temperature cannot be dissipated.

Quantum simulations of the electron dynamics using the
Wigner-Poisson system were performed more recently: as expected,
the Vlasov results were recovered in the large excitation regime
$\delta v > 0.08 v_F$. For smaller excitations, a different regime
appears, in which the ballistic oscillations described above are
no longer observed. Further work is in progress on this issue
\cite{Jasiak}.

\subsubsection{Beyond the mean field} The mean-field approach
described above is appropriate to describe the electron dynamics
on very short time scales ($<100\rm fs$). On a longer time scale
(0.1--1ps), the injected energy is redistributed among the
electrons via electron-electron (e-e) collisions. Electron-phonon
(e-ph) thermalization (i.e., coupling to the ionic lattice) is
generally supposed to occur on even longer time scales. However,
the results of Refs. \cite{Sun,Fann} on thin gold films have shown
that nonequilibrium electrons start interacting with the lattice
earlier than expected, so that a clear-cut separation between e-e
and e-ph relaxation is not entirely pertinent.

The phase-space approach is particularly well-suited to include
corrections that go beyond the mean-field picture. This can be
done with relative ease for semiclassical models (Vlasov), by
using a Boltzmann-like e-e collision integral that respects
Pauli's exclusion principle ({\"U}hling-Uhlenbeck model) \cite{UU}:
\be\left(\frac{\partial f}{\partial t} \right)_{\rm UU}=\int
\frac{d^3 {\bf p_2} d\Omega}{(2\pi\hbar)^3}
~\sigma(\Omega)|v_{12}| (f_1 f_2 \overline{f}_3 \overline{f}_4 -
f_3 f_4 \overline{f}_1 \overline{f}_2)~, \label{vuu} \ee
where $v_{12}$ is the relative velocity of the colliding particles
1 and 2, $\sigma(\Omega)$ is the differential cross section
depending on the scattering angle $\Omega$, and indices 3 and 4
label the outgoing momenta, $f_i = f(\vec{r}, \vec{p}_i , t)$ and
$\overline{f}_i = 1-f_i/2$. This collision term is similar to the
well known classical Boltzmann collision term but for Pauli
blocking factors $\overline{f}_i \overline{f}_j$. As known from
solid-state physics, this blocking factor plays a dramatic role
for electronic systems \cite{Ashcroft}. At $T_e = 0 \rm K$, all
collisions are Pauli blocked and the collisional mean-free path of
the electrons becomes infinite. But if the system becomes excited,
phase space opens up and activates the collision term. The effect
of the above e-e collision term on the semiclassical Vlasov
dynamics in metal clusters was investigated numerically in
\cite{Domps}.

It is conceptually harder to include collisions in fully quantum
models. A significant constraint is that nonunitary corrections to
the Wigner equation should be written in `Lindblad form'
\cite{Lindblad}, which guarantees that the evolved Wigner function
corresponds to a positive-definite density matrix.

The {\"U}hling-Uhlenbeck collision term (\ref{vuu}) is a complicated
nonlinear integral, which is difficult to implement in a numerical
code. It is therefore useful to construct some simplified
collision terms that are more easily amenable to numerical
treatment. In the following, we briefly illustrate two simple
models of e-e and e-ph collisions that we have employed in our
previous works.

\begin{figure}
\centering
\includegraphics[height=4.cm,width=7cm]{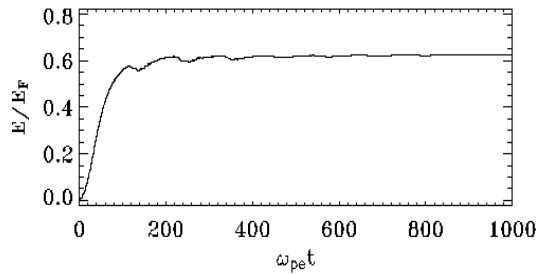}
\caption{Evolution of the thermal energy for a case with e-e
collisions and $L=100 L_F \simeq 120\rm\AA$.} \label{fig:eth-coll}
\end{figure}

\paragraph{Electron-electron collisions.} To model e-e
collisions, a relaxation term is added to the right-hand side of
the Vlasov or Wigner equation:
\be \left(\frac{\partial f}{\partial t} \right)_{\rm e-e} \equiv
-\nu_{ee}(T_e)(f-f_{\infty}), \ee
where $\nu_{ee}$ is the average e-e collision rate and
$f_{\infty}(x,v)$ is a Fermi-Dirac distribution. The idea behind
this model is that the electron distribution will eventually
relax, on a time scale of the order $\nu_{ee}^{-1}$, towards a
Fermi-Dirac equilibrium $f_{\infty}$ with total energy equal to
that of the initial electron distribution $f(x,v,t=0^+)$,
including of course the initial excitation energy. For electrons
near the Fermi surface, the e-e collision rate can be written as
\cite{Pines}
\be \nu_{ee}(T_e) = a (k_B T_e)^2, \label{nuee} \ee
where $a$ is a (dimensional) proportionality constant. The latter
has been estimated from numerical simulations of the electron
dynamics in sodium clusters \cite{Domps}, yielding $a \simeq
0.4~{\rm fs^{-1}eV^{-2}}$, which is also compatible with the
analytical prediction given by the random phase approximation
\cite{Pines}. The electron temperature is computed instantaneously
during the simulation, and plugged into the expression for the
collision rate (\ref{nuee}). It is important to underline that the
above model for e-e collisions, though simple, is completely
self-contained and requires no additional {\it ad-hoc} parameters.
The model has been applied to the electron dynamics in thin metal
films. The slow ballistic oscillations of Fig. \ref{fig:film} are
still observed, although they are damped on a time scale of the
order of $500\omega_{pe}^{-1} \simeq 50\rm fs$ (see Fig.
\ref{fig:eth-coll}).

\paragraph{Electron-phonon collisions.} By coupling to the ionic
lattice, the electrons progressively relax to a thermal
distribution with a temperature equal to that of the lattice
$T_i$. This relaxation time is generally termed $\tau_1$ in the
semiconductor literature. In addition, the lattice acts as an
external environment for the electrons, leading to a loss of
quantum coherence over a time scale $\tau_2$ (decoherence time).
The relaxation and decoherence times correspond, respectively, to
the decay of diagonal and nondiagonal terms in the density matrix
describing the electron population.

Such environment-induced decoherence can be modeled, in the Wigner
representation, by an appropriate friction-diffusion term
\cite{Zurek}:
\begin{equation}
\label{scatt} \left(\frac{\partial{f}}{\partial{t}}\right)_{\rm
e-ph} = 2\gamma \frac{\partial{(vf)}}{\partial{v}} + D_v
\frac{\partial^2{f}}{\partial{v^2}}+ D_x
\frac{\partial^2{f}}{\partial{x^2}}~,
\end{equation}
where $\gamma$ is the relaxation rate (inverse of the relaxation
time $\tau_1$), and $D_v$, $D_x$ are diffusion coefficients in
velocity and real space respectively, which are related to the
decoherence time $\tau_2$ and depend on the lattice temperature
$T_i$. The effect of the diffusive terms is to smooth out the fine
structure of the Wigner function, thus suppressing interference
phenomena, which are a typically quantum effect. Finally, we
recall that, in order to preserve the positivity of the density
matrix associated to the Wigner distribution function, the e-ph
collision term (\ref{scatt}) must be in Lindblad form
\cite{Lindblad}. This is automatically achieved \cite{Isar} if the
coefficients respect the inequality $D_v D_x \geq \gamma^2
\hbar^2/4m^2$.

\subsection{Hydrodynamical models: from micro to macro}
\label{sec:hydro}
Despite its considerable interest, the Wigner-Poisson (WP)
formulation presents some intrinsic drawbacks : (i) it is a
nonlocal, integro-differential system; and (ii) its numerical
treatment requires the meshing of the whole phase space. Moreover,
as is often the case with kinetic models, the Wigner-Poisson
system gives more information than one is really interested in.
For these reasons, it would be useful to obtain an accurate
reduced model which, though not providing the same detailed
information, is still able to reproduce the main features of the
physical system under consideration.

In this section, we will derive an effective Schr\"odinger-Poisson
(SP) system, which, in an appropriate limit, reproduces the
results of the kinetic WP formulation \cite{qfluid}. In order to
obtain the effective SP system, we will first derive a system of
reduced hydrodynamic (or fluid) equations by taking moments of the
WP system. It will be shown that the pressure term appearing in
the fluid equations can be decomposed into a classical and a
quantum part. With some reasonable hypotheses on the pressure
term, the fluid system can be closed. For simplicity of notation,
only one-dimensional problems will be considered, but the results
can be easily extended to higher dimensions.

In order to derive a fluid model, we take moments of Eq.
(\ref{wignereq}) by integrating over velocity space. Introducing
the standard definitions of density, mean velocity, and pressure
\begin{equation}
\label{fluid} n = \int f\,dv \,, \quad u = \frac{1}{n}\int fv\,dv
\,, \quad P = m\left(\int fv^{2}dv - nu^{2}\right) \, ,
\end{equation}
it is obtained
\begin{eqnarray}
\label{cont} \frac{\partial\,n}{\partial\,t} &+&
\frac{\partial\,(nu)}{\partial\,x} = 0 \,, \\
\label{force} \frac{\partial\,u}{\partial\,t} &+&
u\frac{\partial\,u}{\partial\,x} =
\frac{e}{m}\frac{\partial\,\phi}{\partial\,x} - \frac{1}{mn}
\frac{\partial\,P}{\partial\,x} .
\end{eqnarray}
We immediately notice that, surprisingly, Eqs.
(\ref{cont})-(\ref{force}) do not differ from the ordinary
evolution equations for a classical fluid. It can be shown,
however, that quantum effects are actually hidden in the pressure
term, which may be decomposed into a classical and a quantum part.

By using the definition of the Wigner function (\ref{wigfunc}) and
representing each state in terms of its amplitude
$\sqrt{n_{\alpha}}$ and phase $S_\alpha$
\begin{equation}
\label{psi} \psi_{\alpha}(x,t) =
\sqrt{n_{\alpha}(x,t)}\exp{(iS_{\alpha}(x,t)/\hbar)},
\end{equation}
we obtain that $P = P^{C} + P^{Q}$. The classical part of the
pressure can be written as
\be P^C = mn \left[\sum_{\alpha} p_\alpha \frac{n_\alpha}{n}
u^2_\alpha - \left(\sum_\alpha p_\alpha \frac{n_\alpha}{n}
u_\alpha\right)^2 \right] \equiv mn (\langle u_\alpha^2 \rangle -
\langle u_\alpha \rangle^2), \ee
where $m u_\alpha = \partial S_\alpha/\partial x$ [the
$u_\alpha$'s should not be mistaken with the {\it global} mean
velocity $u$ defined in Eq. (\ref{fluid})]. This is the standard
expression for the pressure as velocity dispersion, thus
justifying the term `classical' pressure.

The quantum part of the pressure is written as
\begin{eqnarray}
P^Q &=&
\frac{\hbar^2}{2m}\sum_{\alpha}p_{\alpha}\left(\left(\frac{
\partial\,\sqrt{n_\alpha}}{\partial\,x}\right)^2 - \sqrt{n_\alpha}\frac{\partial^{2}\sqrt{n_\alpha}}
{\partial\,x^2}\right).
\end{eqnarray}
It can be shown that, for distances larger that the Thomas-Fermi
screening length $L_F$, one can replace $n_\alpha$ with $n$, the
total density as defined in Eq. (\ref{fluid}). In order to close
the fluid system (\ref{cont})-(\ref{force}) one still has to
express the classical pressure in terms of the density $n$. This
is the standard procedure adopted in classical hydrodynamics: the
relation $P^C(n)$ is the equation of state, and depends on the
particular conditions of the system, notably its temperature.

With these hypotheses, the Eq. $(\ref{force})$ reduces to
\begin{equation}
\label{force2} \frac{\partial\,u}{\partial\,t} +
u\frac{\partial\,u}{\partial\,x} =
\frac{e}{m}\frac{\partial\,\phi}{\partial\,x} -
\frac{1}{m}\frac{\partial\,W}{\partial\,x} +\frac{\hbar^2}{2m^2}
\frac{\partial}{\partial
x}\left(\frac{\partial^{2}(\sqrt{n})/\partial\,x^2}
{\sqrt{n}}\right) \,,
\end{equation}
where we have defined the effective potential
\begin{equation}
\label{efpot} W(n) = \int^{n}\frac{dn'}{n'}\frac{dP^{C}(n')}{dn'}.
\end{equation}
Equations (\ref{cont}) and (\ref{force2}) constitute the quantum
hydrodynamical approximation to the full Wigner (or Hartree)
equation.

It is now possible to combine Eqs. (\ref{cont}) and (\ref{force2})
into an effective nonlinear Schr\"odinger equation. To this
purpose, let us define the effective wavefunction
\begin{equation}
\Psi = \sqrt{n(x,t)}\exp{(iS(x,t)/\hbar)} \,,
\end{equation}
with $S(x,t)$ defined according to $m u(x,t) =
\partial\,S(x,t)/\partial\,x$. We obtain that $\Psi(x,t)$
satisfies the equation
\begin{equation}
\label{nlse} i\hbar\frac{\partial\Psi}{\partial\,t} = -
\,\frac{\hbar^2}{2m}\frac{\partial^{2}\Psi}{\partial\,x^2} -
e\phi\Psi + W\Psi \, .
\end{equation}

By linearizing Eqs. (\ref{cont}) and (\ref{force2}) around a
homogeneous equilibrium, we obtain the following dispersion
relation
\begin{equation}
\label{dispfluid} \omega^2 = \omega_p^{2} + {v_0}^2 k^{2} +
\frac{\hbar^2 k^4}{4 m^2},
\end{equation}
where $mv_0^2=(dP^C/dn)_{n=n_0}$. It can be proven that, by an
appropriate choice of the equation of state $P^C(n)$, Eq.
(\ref{dispfluid}) reproduces correctly the leading terms of the
Hartree or Wigner dispersion relation.

To summarize, we have shown that, under appropriate conditions,
the Hartree or Wigner models can be reduced to a set of two
hydrodynamical equations (\ref{cont}) and (\ref{force2}), or,
equivalently, to a single nonlinear Schr\"odinger equation
(\ref{nlse}). The two hypotheses used for this reduction were that
(i) all quantities vary on a length scale larger than $L_F$; and
(ii) the equation of state for the classical pressure is $P^C =
P^{C}(n)$ (standard fluid closure).

\begin{figure}
\centering
\includegraphics[height=4.5cm]{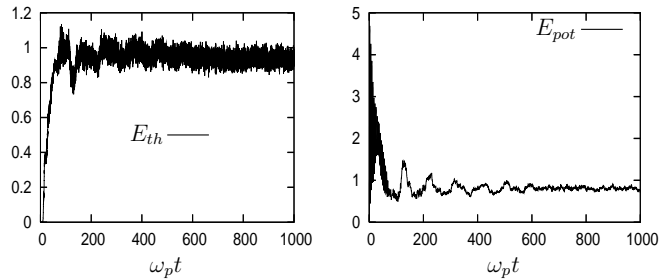}
\caption{Time evolution of the
thermal and potential energies (normalized to $E_F$) of the
electron population, obtained using a quantum hydrodynamics
model.} \label{fig:fluid}
\end{figure}

\paragraph{Example --- Thin metal films}
We have studied the electron dynamics in a thin metal film using
the above quantum hydrodynamical model \cite{Crouseilles}. A
preliminary result is shown in Fig. \ref{fig:fluid}, where we plot
the evolution of the thermal and potential energies against time.
In order to compare to the Vlasov simulations described in Sec.
\ref{sec:phase-space}, the hydrodynamic equations are solved in
the semiclassical limit, i.e. using a small value of the Planck
constant normalized to $E_F/\omega_{p}$ (note however that here
the initial excitation $\delta v = 0.22v_F$ is larger compared to
the case of Fig. \ref{fig:film}, where $\delta v = 0.08v_F$). The
hydrodynamic results display some coherent oscillations at high
frequency, which are a typical signature of quantum effects.
Nevertheless, the initial increase of the thermal energy is
clearly captured and the subsequent ballistic oscillations are
still visible, particularly on the potential energy.

\section{Spin dynamics} \label{sec:spin}
The dynamics of magneto-optical processes in metallic
nanostructures depends on the temporal and spatial characteristics
that are being investigated. Short time scale ($t < 10^{-12}$ s)
have only been explored recently. In 1996, the group of Jean-Yves
Bigot in Strasbourg highlighted the existence of ultrafast
demagnetization processes (within less than a hundred
femtoseconds) induced by femtosecond laser pulses in ferromagnetic
thin films \cite{demagnet}. These demagnetization processes are
not yet fully understood.

From a theoretical point of view, very little is known on the
time-dependent magneto-optical response of metallic nanostructures
to an ultrafast optical pulse. The main difficulty is to provide
an adequate description of the interplay between electronic and
spin degrees of freedom in the metal. So far, only two theoretical
models have been proposed to explain this effect
\cite{Zhang,Koopmans}. These works are based on two different
mechanisms: in \cite{Zhang}, the spin-orbit coupling is invoked,
whereas in \cite{Koopmans} phonon or impurity mediated spin-flip
scattering is privileged. Unfortunately, the parameters employed
in \cite{Zhang} are not realistic and the model developed in
\cite{Koopmans} is a phenomenological approach that does not allow
quantitative predictions. From the above considerations it follows
that there is a need for the development of efficient theoretical
models able to explain in a quantitative manner the experimental
findings.

A proper treatment of spin dynamics requires an extension of our
model (TDLDA) to include spin degrees of freedom. In the
following, the formalism of the time-dependent local-spin-density
approximation (TDLSDA) in the linear regime (including also its
extension to finite temperature) is presented. A second part will
be devoted to the nonlinear dynamics.

\subsection{Linear response: local spin density approximation}
\label{sec:tdlsda}
The generalization of the linear TDLDA to spin polarized electron
systems has been performed by Rajagopal \cite{Rajagopal}. In the
following we provide the basic equations of this approach
including its extension to finite temperature.

Within the framework of DFT one can calculate the spin density
matrix $n_{\sigma \sigma'}(\vec{r})$ defined as
\begin{equation}
n_{\sigma \sigma'}(\vec{r})=\langle 0
|\hat{\psi}^{+}_{\sigma}(\vec{r}) \hat{\psi}^{}_{\sigma'}(\vec{r})
|0\rangle
\end{equation}
where $\hat{\psi}^{+}_{\sigma}(\vec{r})$ and
$\hat{\psi}^{}_{\sigma}(\vec{r})$ are the wave field operators
corresponding to the creation and annihilation of an electron with
spin $\sigma$ at position $r$, and $|0\rangle$ is the ground state
of the system. When the system is subjected to a small local
spin-dependent external potential $\delta V^{\sigma
\sigma'}_{ext}(\vec{r};\omega)$ (this quantity describes the
coupling of the charge and spin of the electrons to external
electric and magnetic fields) the spin-density response function
is defined through the equation
\begin{equation}
\delta n_{\sigma \sigma'} (\vec{r};\omega )=\sum_{\sigma_1
\sigma_2}\int \chi_{\sigma \sigma', \sigma_1 \sigma_2}
(\vec{r},\vec{r}^{\prime };\omega )\ \delta V^{\sigma_1
\sigma_2}_{ext}(\vec{r}^{\prime };\omega )\ d\vec{r}^{\prime }\;.
\label{resp_ss}
\end{equation}

For the sake of simplicity, we restrict ourself to the case of
collinear magnetism, i.e. to the case of a uniform direction of
magnetization. This restriction leads to a diagonal spin-density
matrix ($n_{\sigma \sigma'}= n_{\sigma} \delta_{\sigma \sigma'}$)
and simplified expressions. The spin-density response function
defined in Eq. (\ref{resp_ss}) reduces to
\begin{equation}
\delta n_{\sigma} (\vec{r};\omega )=\sum_{\sigma'}\int
\chi_{\sigma \sigma'} (\vec{r},\vec{r}^{\prime };\omega )\ \delta
V^{\sigma'}_{ext}(\vec{r}^{\prime };\omega )\ d\vec{r}^{\prime }
\end{equation}
which can be rewritten as
\begin{equation}
\delta n_{\sigma} (\vec{r};\omega )=\sum_{\sigma'}\int
\chi^0_{\sigma \sigma'} (\vec{r},\vec{r}^{\prime };\omega )\
\delta V^{\sigma'}_{eff}(\vec{r}^{\prime };\omega )\
d\vec{r}^{\prime }
\end{equation}
with
\begin{eqnarray}
\delta V^{\sigma}_{eff}(\vec{r};\omega ) &=&  \delta
V_{ext}^{\sigma}(\vec{r};\omega ) \nonumber \\ &+& \sum_{\sigma'}
\int \left\{\frac{e^2/4\pi\epsilon_0}{\left| \vec{r}-\vec{r}%
^{\prime }\right|} + f_{xc}^{\sigma
\sigma'}(\vec{r},\vec{r}^{\prime };\omega )\right\}\delta
n_{\sigma'}(\vec{r}^{\prime };\omega ) \; d\vec{r}^{\prime }\;.
\end{eqnarray}
In the above expression the function $f_{xc}^{\sigma
\sigma'}(\vec{r},\vec{r}^{\prime };\omega )$ is the Fourier
transform of the time-dependent kernel defined by $f_{xc}^{\sigma
\sigma'}(\vec{r},t;\vec{r}^{\prime },t^{\prime })\equiv \delta
V_{xc }^{\sigma}(\vec{r},t)/\delta n_{\sigma'} (\vec{r}^{\prime
},t^{\prime })$ and $\chi ^{0}_{\sigma
\sigma'}(\vec{r},\vec{r}^{\prime };\omega )\ $ is the
non-interacting retarded spin-density correlation function. For
spin polarized electron systems the exchange-correlation potential
is defined as
\begin{equation}
V_{xc}^{\sigma}(\vec{r})=\left[\frac{\partial} {\partial
n_{\sigma}} \left\{{n}\omega_{xc}(n_{+},n_{-})\right\}
\right]_{n_{+}=n_{+}(\vec{r});n_{-}=n_{-}(\vec{r})} \;,
\label{easxc}
\end{equation}
where $\Omega_{xc}[n_{+},n_{-}]=\int n(\vec{r})
\omega_{xc}\left(n_{+}(\vec{r}),n_{-}(\vec{r})\right) d\vec{r}$ is
the exchange-correlation thermodynamic potential and $\omega_{xc}$
the exchange-correlation thermodynamic potential per particle of
the homogeneous electron gas calculated at the local density $n$
and magnetization $m=n_+-n_-$. By noting that
\[\frac{\partial}{\partial n_{\sigma}} \left\{
n\omega_{xc}(n_+,n_-)\right\}= \frac{\partial}{\partial n} \left\{
n\omega_{xc}(n,m)\right\}+\sigma \frac{\partial}{\partial m}
\left\{ n\omega_{xc}(n,m)\right\},\]
the expression (\ref{easxc}) can be rewritten as \cite{Katsnelson}
\begin{equation}
V_{xc}^{\sigma}(\vec{r})=\left[\frac{\partial} {\partial n}
\left\{{n}\omega_{xc}(n,m)\right\}
\right]_{n=n(\vec{r});m=m(\vec{r})} + \sigma\mu_{B}
B_{xc}(\vec{r}) \;, \label{Bxc}
\end{equation}
where $B_{xc}(\vec{r})=\mu_B^{-1} \left[\frac{\partial} {\partial
m} \left\{{n}\omega_{xc}(n,m)\right\}
\right]_{n=n(\vec{r});m=m(\vec{r})}$ is the exchange-correlation
magnetic field acting on spin, and $\mu_B=e\hbar/(2m)$ is the Bohr
magneton. This is an \textit{internal} magnetic field. The
response functions $\chi ^{0}$ and $\chi$ are related by an
integral equation (to be more precise, due to the spin degree of
freedom, it is a matrix integral equation)
\begin{eqnarray}
\chi_{\sigma \sigma'}(\vec{r},\vec{r}^{\prime };\omega ) &=&\chi
^{0}_{\sigma
\sigma'}(\vec{r},\vec{r}%
^{\prime };\omega )+\sum_{\sigma_1 \sigma_2}\int \int \chi_{\sigma
\sigma_1}^{0}(\vec{r},\vec{r}^{\prime \prime
};\omega )  \nonumber \\
&\times &\ K^{\sigma_1 \sigma_2}(\vec{r}^{\prime \prime
},\vec{r}^{\prime \prime \prime };\omega )\ \chi_{\sigma_2
\sigma'}(\vec{r}^{\prime \prime \prime },\vec{r}^{\prime };\omega )\ d\vec{r}%
^{\prime \prime }d\vec{r}^{\prime \prime \prime }, \label{Dyson-s}
\end{eqnarray}
with the residual interaction defined by
\begin{equation}
K^{\sigma_1 \sigma_2}(\vec{r},\vec{r}^{\prime };\omega
)=\frac{e^2}{4\pi\epsilon_0 |\vec{r}-\vec{r}^{\prime
}|}\delta_{\sigma_1\sigma_2}+f_{xc}^{\sigma_1
\sigma_2}(\vec{r},\vec{r}^{\prime };\omega ). \label{residual-s}
\end{equation}
As for TDLDA, in the {\it adiabatic} local-density approximation
(ALDA) the exchange-correlation kernel is frequency-independent
and local and reduces to
\begin{equation}
f_{xc}^{\sigma \sigma'} (\vec{r},\vec{r}^{\prime }) =\left[
\frac{\partial^2 [n\omega_{xc} (n,m)] }{\partial n_{\sigma}
\partial n_{\sigma'}}
 \right]_{n=n(\vec{r});m=m(\vec{r})}\delta
\left( \vec{r}-\vec{r}^{\prime }\right) \label{fxc}\;.
\end{equation}
It should be mentioned that the functional $\omega_{xc}$ in the
above expression should be the same as the one used in the
calculation of the ground state (see Eq. (\ref{easxc})). By using
the same field-theory techniques employed previously for TDLDA
(see Sec. \ref{sec:DFT}), one can show that the free response
function reads
\begin{eqnarray}
\chi ^{0}_{\sigma \sigma'}(\vec{r},\vec{r}^{\prime };\omega
;T_{e}) &=&\delta_{\sigma \sigma'}\sum_{k}\ f_{k}^{\sigma}\
\phi_{k}^{\sigma *}(\vec{r})\phi _{k}^{\sigma}(\vec{r}^{\prime })\
G_{+}^{\sigma}(\vec{r},\vec{%
r}^{\prime };\varepsilon_{k}^{\sigma}+\hbar \omega ;T_{e})  \nonumber \\
&+&\sum_{k}f_{k}^{\sigma}\ \phi _{k}^{\sigma}(\vec{r})\phi
_{k}^{\sigma *}(\vec{r}^{\prime })\ G_+^{\sigma
*}(\vec{r},\vec{r}^{\prime };\varepsilon _{k}^{\sigma}-\hbar
\omega ;T_{e})\;,
\end{eqnarray}
where $\phi_{k}^{\sigma}(\vec{r})$ and $\varepsilon _{k}^{\sigma}$
are the one-electron Kohn-Sham wave functions and energies,
respectively. $G_{+}^{\sigma}$ is the one-particle retarded
Green's function for the spins $\sigma$ and $f_{k}^{\sigma}=\left[
1+\exp \left\{ (\varepsilon _{k}^{\sigma}-\mu )/k_{B}T_{e}\right\}
\right] ^{-1}$. Similarly to TDLDA, we have assumed that the
residual interaction (\ref{residual-s}) is temperature
independent. Thus, it is consistent with the use of
$\omega_{xc}(n,m)=\epsilon_{xc}(n,m)$ in the calculation of the
ground-state properties.

From the above formalism one can compute the dipolar absorption
cross-section
\begin{equation} \sigma \left( \omega ;T_{e}\right)
=\frac{\omega} {\varepsilon_{0} c} \mathop{\rm Im} \left[ \alpha
\left( \omega ;T_{e}\right) \right],
\end{equation}
where $\alpha$ is the frequency-dependent dipole {\em electric}
polarizability defined as
\begin{eqnarray}
\alpha \left( \omega ;T_{e}\right) &=&\int \left[\delta n_+(\vec{r};\omega; T_e )
+\delta n_{-}(\vec{r};\omega; T_e )\right]\ \delta V_{%
ext}(\vec{r};\omega )\ d\vec{r} \;.
\end{eqnarray}
By analogy, one defines a quantity which is constructed from the
local magnetization (instead of the local density)
\begin{equation} \sigma_m \left( \omega ;T_{e}\right)
=\frac{\omega} {\varepsilon_{0} c} \mathop{\rm Im} \left[ \alpha_m
\left( \omega ;T_{e}\right) \right],
\end{equation}
where $\alpha_m$ is the frequency-dependent dipole {\em magnetic}
polarizability defined as
\begin{eqnarray}
\alpha_m \left( \omega ;T_{e}\right) &=&\int \left[\delta n_+(\vec{r};\omega; T_e )
-\delta n_{-}(\vec{r};\omega; T_e )\right]\ \delta V_{%
ext}(\vec{r};\omega )\ d\vec{r} \;.
\end{eqnarray}
On can show that $\sigma_m$ fulfils the following sum rule
\begin{equation}
\int \sigma_m \left( \omega ;T_{e}\right) d \omega= \frac{2\pi^2
M(T_e)}{c}\;
\end{equation}
where $M=N^{+}-N^{-}$ is the total magnetization of the system
($N^{+}$ being the number of spins up and $N^{-}$ the number of
spins down). It is worth mentioning that $M$ is generally
temperature dependent \cite{Maurat}.

\subsection{Nonlinear response: Phase-space methods}
\label{sec:spin-wigner} In order to investigate the nonlinear
regime of the charge and spin dynamics, a phase-space approach is
particularly interesting. In this paragraph, we will construct a
Wigner equation that includes spin effects in the local density
approximation, and show that its classical limit takes the form of
a Vlasov equation.

The starting point for the derivation are the time-dependent
Kohn-Sham (KS) equations described in Sec. \ref{sec:tdlsda}. In
terms of the Pauli 2-spinors
\begin{eqnarray*}
 {\Psi}_i (\bf{r}, t) & = & \left(\begin{array}{c}
    \Psi_i^{\uparrow} (\bf{r}, t)\\
    \Psi_i^{\downarrow} (\bf{r}, t)
  \end{array}\right)
\end{eqnarray*}
the KS equations can be written as:
\begin{eqnarray}
    i \hbar \frac{\partial \Psi_i}{\partial t} & = &
    \left[\left( - \frac{\hbar^2}{2 m} \nabla^2 + V ({\bf r}, t)\right)
    {\bf I}+ \mu_B \vec{\sigma}
    \cdot \vec{B} ({\bf r}, t)\right] {\Psi_i}({\bf r}, t)
    \label{eq:ks-spin}
\end{eqnarray}
where $V({\bf r}, t) = V_{\rm {ext}} ({\bf r}, t) + V_H({\bf r},
t) + V_{\rm{xc}}^0 ({\bf r}, t)$, $\mu_B$ is Bohr's magneton,
$\vec{\sigma}=(\sigma_x,\sigma_y,\sigma_z)$ are the $2\times 2$
Pauli matrices, and ${\bf I}$ is the identity matrix. Here,
$V_{\rm {ext}}$ is an external potential (e.g. ionic jellium,
external electric field, ...), $V_H$ is the Hartree potential that
obeys Poisson's equation, and $V_{\rm{xc}}^0$ is the scalar part
of the exchange-correlation potential. The magnetic field
$\vec{B}=\vec{B}_{\rm ext}+\vec{B}_{\rm xc}$ is composed of an
external part and an `internal' part that stems from the exchange
and correlation energy [see Eq. (\ref{Bxc})]. In the so-called
`collinear' approximation, the latter reduces to $\vec{B}_{\rm
xc}= B_{\rm xc} \hat{z}$.

\subsubsection{Equation of motion for the density matrix}
By defining the density matrix
\begin{eqnarray}
\rho^{\eta \eta'}({\bf r}, {\bf r}') = \sum_{i} \Psi^{\eta}_{i}
({\bf r})  \Psi^{\eta'\star}_{i}({\bf r}') \label{eq:dmatrix}
\end{eqnarray}
where $\eta = \uparrow, \downarrow$, the KS equations
(\ref{eq:ks-spin}) can be written in the following compact form
(Von Neumann equation):
\be
  i \hbar \frac{\partial\rho}{\partial t} =  [H, \rho],
  \label{eq:neumann}
\ee
where
 \be
  \rho =  \left(\begin{array}{cc}
    \rho^{\uparrow \uparrow} & \rho^{\uparrow \downarrow}\\
    \rho^{\downarrow \uparrow} & \rho^{\downarrow \downarrow}
  \end{array}\right) ~;~~~  H = \left(\begin{array}{cc}
    h^{\uparrow \uparrow} & h^{\uparrow \downarrow}\\
    h^{\downarrow \uparrow} & h^{\downarrow \downarrow}
  \end{array}\right).
\ee
The only nondiagonal terms in the Hamiltonian come from the
external or internal magnetic field $\vec{B}$.

We now introduce the following basis transformation for the
Hamiltonian: \be
 H =  h_0 {\bf I} + \vec{h} \cdot \vec{\sigma}
  \label{eq:hvect}
\ee
where $\vec{h} = \left( h_x, h_y, h_z \right)$, and
  \begin{eqnarray}
    h_0 = \frac{h^{\uparrow \uparrow} + h^{\downarrow \downarrow}}{2} & , &
    h_x = \frac{h^{\uparrow \downarrow} + h^{\downarrow \uparrow}}{2}\\
    h_z = \frac{h^{\uparrow \uparrow} - h^{\downarrow \downarrow}}{2} & , &
    h_y = \frac{h^{\downarrow \uparrow} - h^{\uparrow \downarrow}}{2 i}
    \label{eq:hvect1}
  \end{eqnarray}
For the Hamiltonian of Eq. (\ref{eq:ks-spin}), we have
\begin{eqnarray}
  h_0 ({r}) & = & - \frac{\hbar^2}{2 m} \nabla^2 + V ({r},
  t)\\
  h_{\alpha} ({r}) & = & \mu_B B_{\alpha} ({r}, t),~~ \alpha = x,
  y, z
\end{eqnarray}
The same transformation (with identical notation) is also applied
to the density matrix. With these definitions, the equations of
motion for $\rho_0$ and $\rho_{\alpha}$ read as
\begin{eqnarray}
  i \hbar \partial_t \rho_0 &=&  [h_0, \rho_0] + \sum_{\alpha = x, y, z}
  [h_{\alpha}, \rho_{\alpha}]\\
  i \hbar \partial_t \rho_{\alpha} &=& [h_0, \rho_{\alpha}] + [h_{\alpha},
  \rho_0].
\end{eqnarray}

\subsubsection{`Spin' Wigner and Vlasov equations}
By making use of the Wigner transformation
\begin{eqnarray}
  f_0 ({\bf r}, {\bf v}, t) & = & \frac{m}{2 \pi \hbar} \int
  d{\bf \lambda} \rho_0 \left({\bf r}- \frac{{\bf \lambda}}{2},
  {\bf r}+ \frac{{\bf \lambda}}{2}\right) e^{i
  m{\bf v}{\bf \lambda}/ \hbar}\\
  f_{\alpha} ({\bf r}, {\bf v}, t) & = & \frac{m}{2 \pi \hbar} \int
  d{\bf \lambda} \rho_{\alpha} \left({\bf r}-
  \frac{{\bf \lambda}}{2}, {\bf r}+ \frac{{\bf \lambda}}{2}\right)
  e^{i m{\bf v}{\bf \lambda}/ \hbar}
\end{eqnarray}
one can easily obtain the equations of motion for the Wigner
functions:
 \begin{eqnarray*}
    &&\frac{\partial}{\partial t} f_0 +{\bf v} \frac{\partial}{\partial
    {\bf r}} f_0 -\\
    && \frac{m}{2 i \pi \hbar^2} \int d{\bf \lambda}
    \int d{\bf v}' e^{i m ({\bf v}-{\bf v}'){\bf \lambda}/
    \hbar} \left[V \left({\bf r}+ \frac{{\bf \lambda}}{2}\right) - V \left({\bf r}-
    \frac{{\bf \lambda}}{2}\right)\right] f_0 ({\bf r}, {\bf v}', t) - \\
    && \sum_{\alpha} \frac{m\mu_B}{2 i \pi \hbar^2} \int d{\bf \lambda} \int
    d{\bf v}' e^{i m \left({\bf v}-{\bf v}'\right){\bf \lambda}/
    \hbar} \left[B_{\alpha} \left({\bf r}+ \frac{{\bf \lambda}}{2}\right) -
    B_{\alpha} \left({\bf r}- \frac{{\bf \lambda}}{2}\right)\right] f_{\alpha}
    ({\bf r}, {\bf v}', t)  =  0
\end{eqnarray*}

\begin{eqnarray*}
    &&\frac{\partial}{\partial t} f_\alpha +{\bf v} \frac{\partial}{\partial
    {\bf r}} f_\alpha -\\
    && \frac{m}{2 i \pi \hbar^2} \int d{\bf \lambda}
    \int d{\bf v}' e^{i m ({\bf v}-{\bf v}'){\bf \lambda}/
    \hbar} \left[V \left({\bf r}+ \frac{{\bf \lambda}}{2}\right) - V \left({\bf r}-
    \frac{{\bf \lambda}}{2}\right)\right] f_\alpha \left({\bf r}, {\bf v}', t\right) - \\
    && \frac{m\mu_B}{2 i \pi \hbar^2} \int d{\bf \lambda} \int
    d{\bf v}' e^{i m ({\bf v}-{\bf v}'){\bf \lambda}/
    \hbar} \left[B_{\alpha} \left({\bf r}+ \frac{{\bf \lambda}}{2}\right) -
    B_{\alpha} \left({\bf r}- \frac{{\bf \lambda}}{2}\right)\right] f_{0}
    ({\bf r}, {\bf v}', t)  =  0
\end{eqnarray*}
The corresponding Vlasov equations are obtained in the classical
limit $\hbar \to 0$:
\begin{eqnarray}
  \frac{\partial}{\partial t} f_0 +{\bf v} \frac{\partial}{\partial
  {\bf r}} f_0 - \frac{1}{m}  \frac{\partial V}{\partial {\bf r}}
  \frac{\partial f_0}{\partial {\bf v}} - \frac{\mu_B}{m} \sum_{\alpha}
  \frac{\partial B_{\alpha}}{\partial {\bf r}} \frac{\partial
  f_{\alpha}}{\partial {\bf v}} & = & 0\\
  \frac{\partial}{\partial t} f_{\alpha} +{\bf v}
  \frac{\partial}{\partial {\bf r}} f_{\alpha} - \frac{1}{m}
  \frac{\partial V}{\partial {\bf r}} \frac{\partial f_{\alpha}}{\partial
  {\bf v}} - \frac{\mu_B}{m} \frac{\partial B_{\alpha}}{\partial
  {\bf r}} \frac{\partial f_0}{\partial {\bf v}} & = & 0
\end{eqnarray}
with $\alpha = x, y, z$.

Within the collinear approximation, the equations for $\alpha=x,y$
vanish. In this case, it is more convenient revert to the original
representation and use
\begin{eqnarray*}
    f_{\uparrow} & = & f_0 + f_z \\
    f_{\downarrow} & = & f_0 - f_z.
\end{eqnarray*}
The corresponding Vlasov equations then become
\begin{eqnarray}
  \frac{\partial}{\partial t} f_{\uparrow} +{\bf v} \frac{\partial}{\partial
  {\bf r}} f_{\uparrow} - \frac{1}{m} \left( \frac{\partial V}{\partial {\bf r}}
   + \mu_B \frac{\partial B_{z}}{\partial {\bf r}} \right) \frac{\partial
  f_{\uparrow}}{\partial {\bf v}} & = & 0 \\
  \frac{\partial}{\partial t} f_{\downarrow} +{\bf v} \frac{\partial}{\partial
  {\bf r}} f_{\downarrow} - \frac{1}{m}  \left( \frac{\partial V}{\partial {\bf r}}
  - \mu_B \frac{\partial B_{z}}{\partial {\bf r}} \right) \frac{\partial f_{\downarrow}}{\partial {\bf v}} & = &
  0.
\end{eqnarray}

The above Wigner and Vlasov equations can be used to study the
nonlinear spin dynamics in a ferromagnetic nanoparticle or thin
film, using numerical techniques similar to those employed for the
electron dynamics. In their present form, these equations preserve
the total spin, and thus cannot be used to describe the loss of
magnetization observed in experiments \cite{demagnet}. A proper
generalization, along the lines of the e-e and e-ph collision
operators detailed in Sec. \ref{sec:phase-space}, would be
necessary to account for these effects.

\section{Numerical example: the nonlinear many-electron dynamics in an anharmonic quantum well}
\label{sec:results} In order to illustrate qualitatively the
practical implementation of the models described in the previous
sections, we concentrate on a specific -- and relatively simple --
example. We consider an electron population confined in a
one-dimensional anharmonic well defined by the potential
\be V_{\rm conf}(x) = \frac{1}{2}\omega_0^2 m_\star x^2 +
\frac{1}{2}K x^4, \label{eq:v-conf}\ee
where $m_\star$ is the effective electron mass. The frequency
$\omega_0$ can be related to a fictitious homogeneous positive
charge of density $n_0$ via the relation $\omega_0^2 = e^2
n_0/m_\star \varepsilon$. The total potential seen by the
electrons is the sum of the confining potential $ V_{\rm conf}$
and the Hartree potential, which obeys Poisson's equation
\begin{equation}
\label{eq:pois-elec} \frac{\partial^{2}V_H}{\partial\,x^2} =
\frac{e^2}{\varepsilon}\int_{-\infty}^\infty f\,dv ~,
\end{equation}
where $e$ is the absolute electron charge and $\varepsilon$ is the
effective dielectric constant. As initial condition, we take a
Maxwell-Boltzmann distribution with Gaussian density profile
\begin{equation}
f_0(x,v) = \frac{\overline{n}_e}{\sqrt{2\pi k_BT_e/m_\star}
}\exp\left(-\frac{m_\star v^2 +m_\star\omega_0^2 x^2}{2k_B
T_e}\right), \label{ic}
\end{equation}
with temperature $T_e$ and peak density $\overline{n}_e$.

The electron dynamics is mainly determined by two dimensionless
parameters: (i) the `filling fraction' $\eta
=\overline{n}_e/n_0=\omega_p^2/\omega_0^2$, which is a measure of
self-consistent effects (in the limit case $\eta=0$, corresponding
to very dilute electron densities, the Hartree potential is
negligible); and (ii) the normalized Planck constant
$H=\hbar\omega_0/k_B T_e$, which determines the importance of
quantum effects. Notice that a small value of $H$ corresponds to a
large electron temperature.

We use typical parameters for semiconductor quantum wells
\cite{Ullrich,Manfredi-APL}: effective electron mass and
dielectric constant $m_\star=0.067m_e$ and $\varepsilon =
13\varepsilon_0$; volume density $n_0 = 10^{16}{\rm cm}^{-3}$,
oscillator energy $\hbar\omega_0 = 3.98 {\rm meV}$, and oscillator
length $L_{\rm ho} = \sqrt{\hbar/m_\star \omega_0}\simeq 17 \rm
nm$. For $\eta=1$, this yields a maximum surface density for the
electrons $n_s = 4.64 \times 10^{10}{\rm cm}^{-2}$ and a maximum
Fermi temperature $T_F = 29.3{\rm K}$. A low electron temperature
$T_e \simeq 46{\rm K}$ then yields $H\simeq 1$, whereas at room
temperature $T_e \simeq 300{\rm K}$ one has $H\simeq 0.15$.

The electron dynamics is excited by shifting the electron density
of a finite distance $\delta x =L_{\rm ho}$. We will primarily be
interested in the relaxation of the electric dipole, defined as
the center of mass of the electron population: $d(t) = \int \int f
x dxdv / \int \int f dxdv$, and of the average kinetic energy
$E_{\rm kin}= {1\over 2}\int \int f m_\star v^2 dxdv/ \int \int f
dxdv$.

First, we present results obtained from the numerical resolution
of the Wigner equation (\ref{wignereq}), coupled to Poisson's
equation (\ref{eq:pois-elec}). The results were obtained with a
numerical code that combines the split-operator method with fast
Fourier transforms in the velocity coordinate \cite{Suh}. We
explore the electron dynamics for different values of the two
relevant dimensionless parameters, $H$ and $\eta$. The
anharmonicity parameter appearing in the confining potential
(\ref{eq:v-conf}) is fixed to $K=0.1$ (in units where
$\hbar=m_\star=\omega_0=1$). If the confinement were purely
harmonic (i.e., $K=0$), the dipole would simply oscillate at the
frequency $\omega_0$ irrespective of the value of the filling
fraction. This result goes under the name of Kohn's theorem
\cite{Kohn}, and we have checked that it holds for our numerical
simulations. When the confinement is not harmonic, the dipole
should decay because of phase mixing effects.

\begin{figure}
\centering
\includegraphics[height=8cm, width=11cm]{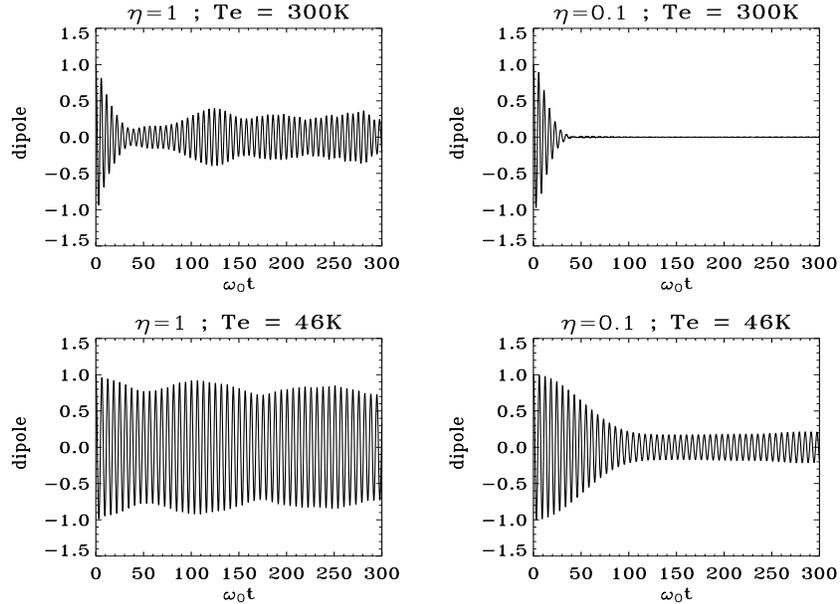}
\caption{Evolution of the electric dipole (in units of $L_{\rm ho}
= 17\rm nm$) obtained from the Wigner-Poisson model, for several
values of $\eta$ and the electron temperature. Time is normalized
to the oscillator frequency.} \label{fig:dipole-wigner}
\end{figure}

\begin{figure}
\centering
\includegraphics[height=8cm, width=11cm]{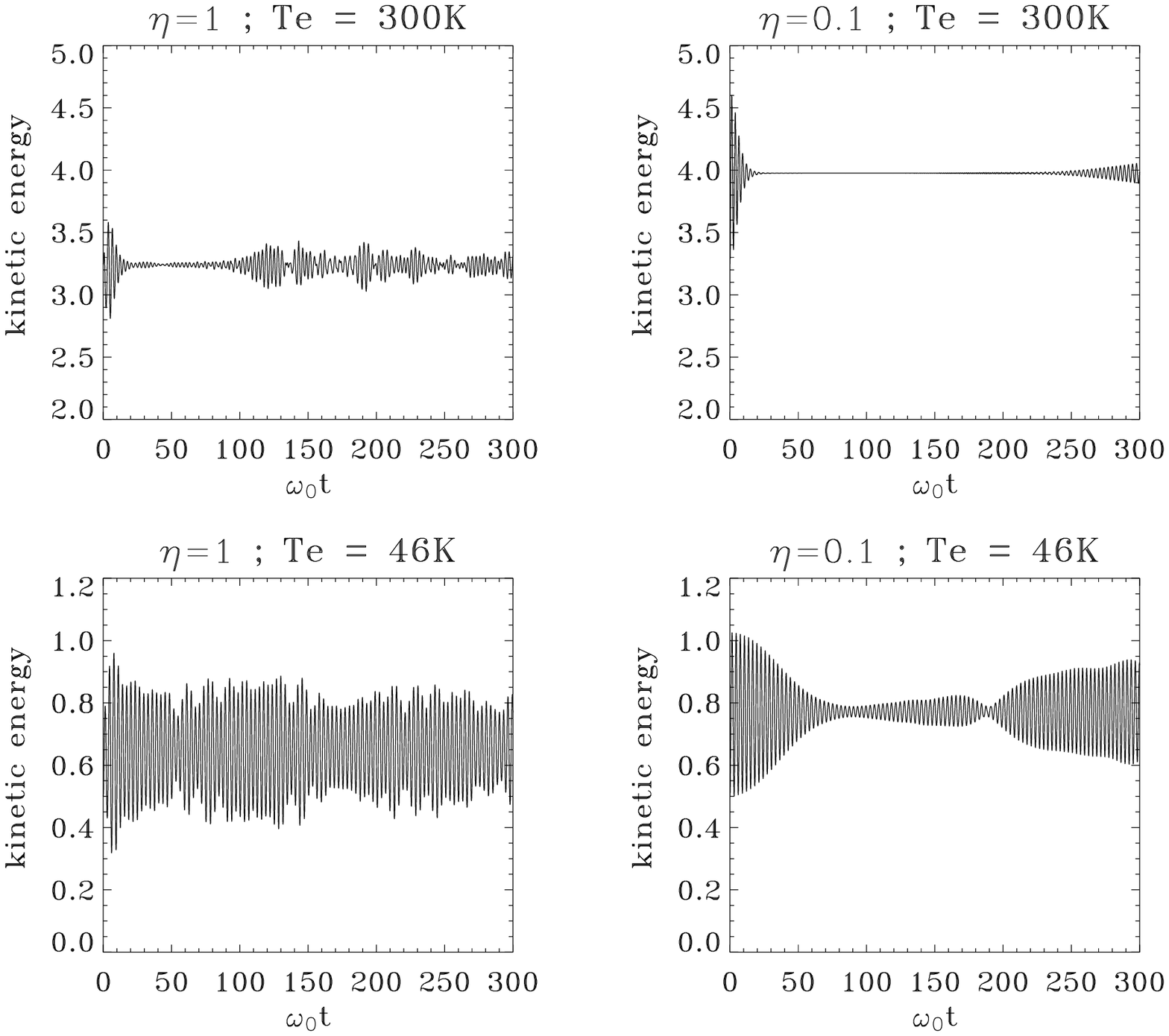}
\caption{Evolution of the kinetic energy (normalized to
$\hbar\omega_0 = 3.98\rm meV$) obtained from the Wigner-Poisson
model, for several values of $\eta$ and the electron temperature.
Time is normalized to the oscillator frequency.}
\label{fig:ekin-wigner}
\end{figure}

The numerical results are shown in Fig. \ref{fig:dipole-wigner}
(dipole) and Fig. \ref{fig:ekin-wigner} (kinetic energy). The fast
oscillations correspond to the center of mass of the electron gas
oscillating in the anharmonic well. For low electron densities and
large temperatures ($\eta=0.1, T_e=300$K), the dipole relaxes to
the bottom of the well, $d\simeq 0$, whereas the kinetic energy
relaxes to a constant asymptotic value. This is a semiclassical
regime where the energy spectrum is almost continuous: the
observed relaxation is due to phase mixing effects.

Decreasing the temperature ($T_e=46$K) while keeping the density
low ($\eta=0.1$) produces a revival that occurs after the kinetic
energy has initially relaxed. This is a typically quantum effect
resulting from the discrete nature of the energy spectrum. The
revival is clearly visible on the kinetic energy, but not so much
on the dipole. When the electron density is large ($\eta=1$),
self-consistent electron-electron interactions (Hartree potential)
prevent the dipole and the kinetic energy from relaxing
completely, even at large temperatures.

Next, we have added a dissipative term to the Wigner equation, in
order to model electron-phonon (e-ph) collisions. This model has
been discussed in Sec. \ref{sec:phase-space}. The relaxation rate
is chosen to be $\gamma=0.001\omega_0$, yielding a realistic
relaxation time $\tau_1 = \gamma^{-1} \simeq 165$ps. The
velocity-space diffusion coefficient is $D_v=\gamma v_{th}$, where
the thermal velocity is $v_{th}=\sqrt{k_BT_e/m_\star}$. The
relaxation time $\tau_2$ depends on the velocity scale: for
instance, a velocity scale $\Delta v$ is damped on a time scale
$\tau_2=\tau_1 \Delta v/v_{th}$. Therefore, for velocity scales
smaller than the thermal velocity, the decoherence time is always
smaller than the relaxation time, in accordance with experimental
findings.

\begin{figure}
\centering
\includegraphics[height=8cm, width=11cm]{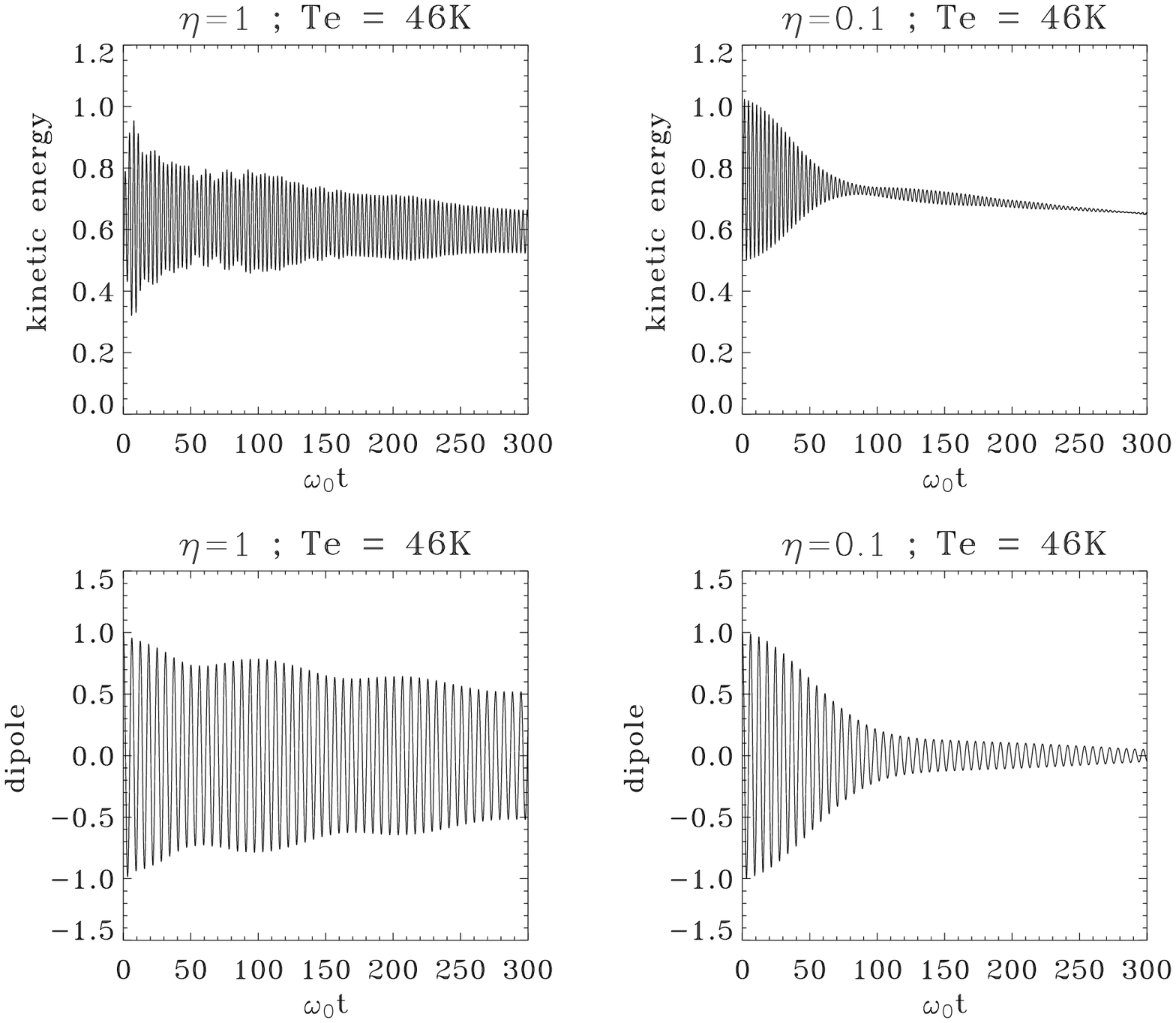}
\caption{Evolution of the kinetic energy (top panels) and electric
dipole (bottom panels), from the Wigner-Poisson model including
e-ph collisions. Same normalizations as in Figs.
\ref{fig:dipole-wigner} and \ref{fig:ekin-wigner}.}
\label{fig:dissip-wigner}
\end{figure}

We simulated the low temperature scenario ($T_e=46$K) in the
presence of e-ph collisions, and observed that the revival
occurring in the kinetic energy for $\eta=0.1$ is now suppressed
(see Fig. \ref{fig:dissip-wigner}). For large densities, however,
the coherence of the electron motion is not lost, and the
relaxation of the dipole and the kinetic energy is only marginally
faster compared to the collisionless regime.

\begin{figure}
\centering
\includegraphics[height=5cm]{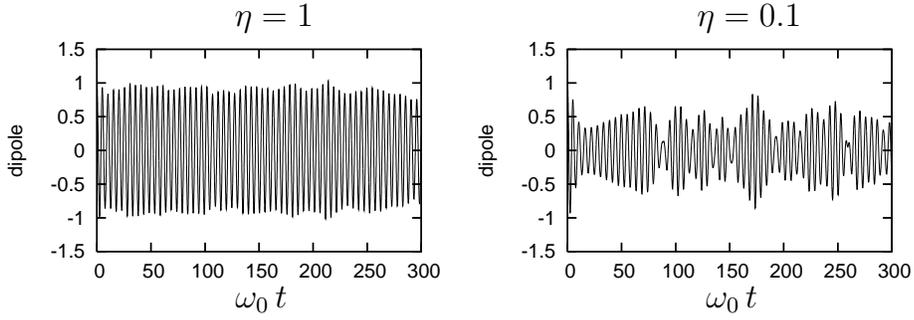}
\caption{Evolution of the electric dipole for $\eta=1$ (left
frame) and $\eta=0.1$ (right frame), obtained from the quantum
hydrodynamic model at $T_e=0$.} \label{fig:dipole-fluid}
\end{figure}

Finally, we want to consider the zero-temperature case. For doing
this, we resort to the hydrodynamical model described in Sec.
\ref{sec:hydro}. The relevant dimensionless parameters now are
$\eta$ and $r_{s0}$, the normalized Wigner-Seitz radius computed
with the background density $n_0$. For $n_0=10^{16}{\rm cm}^{-3}$,
one has $r_{s0} =2.8$. In Fig. \ref{fig:dipole-fluid} we plot the
evolution of the electric dipole for different values of the
filling fraction. Now, even for low electron densities, the dipole
oscillates indefinitely without any appreciable decay. For larger
electron densities, the motion is even more regular. It appears,
therefore, that the dynamics becomes more and more regular as the
electron temperature decreases, i.e. when quantum effect become
more important. As mentioned above, this is essentially due to
phase mixing effect, which become increasingly important in the
semiclassical regime, where the energy levels are almost
continuous.

\section{Conclusions and Perspectives}
\label{sec:conclusions} In this review paper, we have presented
some of the most common theoretical models used to describe the
charge and spin dynamics in metallic and semiconductor
nanostructures. Three levels of description have been identified
(see Fig. \ref{fig:schema}): (i) the full quantum $N$-body
problem, which can only be addressed for small systems by using,
for instance, the Configuration Interaction (CI) method; (ii) mean
field models (Hartree and Wigner) and their generalizations to
include exchange and correlations (Hartree-Fock, density
functional theory); and (iii) quantum hydrodynamical models, which
describe the electron dynamics via a small number of macroscopic
variables, such as the density and the average velocity.

Each of these quantum-mechanical approaches has its classical
counterpart: classical $N$-body models have been developed for
molecular dynamics computations, as well as for gravitational
$N$-body problems; classical mean field models are ubiquitous in
plasma physics (Vlasov-Maxwell equations) and in the study of
self-gravitating objects such as star clusters, galaxies, or even
the entire universe; classical hydrodynamics hardly needs
mentioning, as it is in itself an extremely wide field of
research.

For each approach, we have stressed the difference between the
linear and the nonlinear response. The former is valid for weak
excitations and presupposes that the response is directly
proportional to the excitation. Linear response theory is
generally represented in the frequency domain. In contrast,
nonlinear effects kick in for large excitations, and are best
described in the time domain (this is because the time-frequency
Fourier transform is a linear operation, thus not adapted to
describe nonlinear relations). Although a vast literature on the
linear electronic response is available and dates back from the
works of Drude in the early twentieth century, nonlinear effects
have only been investigated in the last two decades, mainly with
computer simulations.

The mean field level of description is perhaps the most widely
used, as it incorporates, at least to lowest order, some of the
features of the $N$-body dynamics, but still avoids the formidable
complexity of the full problem. A particularly challenging open
problem is the inclusion of dynamical correlations within
mean-field models. Dynamical correlations differ from the
correlations that are included in time-dependent density
functional theory (TDDFT), inasmuch as they cannot be described by
a slowly-varying density functional, as is done in ALDA (adiabatic
local-density approximation). Whereas adiabatic correlations are
described within an essentially Hamiltonian formulation and thus
cannot model irreversible effects, dynamical correlations are
responsible for the relaxation of the electron gas towards
thermodynamical equilibrium. Some recent results have been
obtained using a generalization of TDDFT that relies on the
electron {\em current} as well as the electron density
\cite{Dagosta-Vignale}. The phase-space approach, via the Wigner
formulation, also appears promising to model effects beyond the
mean field, as we have illustrated in Sec. \ref{sec:phase-space}.

Another important issue, which was not mentioned earlier in this
review, is the inclusion of relativistic corrections in the above
models for the electron dynamics. Spin-orbit coupling (which is an
effect appearing at second order in $v/c$) is sometimes taken into
account in a semi-phenomenological way within the Pauli equation.
However, other terms occurring at the same order are often
neglected without further justification. A consistent derivation
of relativistic effects to a certain order in $v/c$ can of course
be carried out, starting from the Dirac equation, for the case of
a single particle in an external electromagnetic field
\cite{Landau2}. For a many-body system, this issue is much
trickier and is the object of current investigations.

Nanostructures are by definition finite-size objects. Due to the
presence of boundaries and interfaces, the electron dynamics can
thus display novel and unexpected features compared to bulk
matter. For example, as the elastic and inelastic scattering
length ($\sim 10-50$nm for bulk metals) are much longer than the
size of the system, an electron -- or a group of electrons -- can
travel coherently through the length of the system, thus leading
to ballistic transport between the surfaces. The theoretical tools
to study finite-size nano-objects are also relatively recent, and
have been developed alongside the experimental breakthroughs that
made these objects widely available.

If the electron dynamics in nanosized objects has received
considerable attention for the last thirty years, the {\em spin}
dynamics is a much younger field of research, both experimentally
and theoretically. Nevertheless, the already existing applications
to memory storage and processing, and the still speculative, but
highly enthralling, developments in quantum computing, have
stimulated a large number of works in this direction. In Sec.
\ref{sec:spin} we have illustrated how the models for the electron
dynamics can be extended to include the spin degrees of freedom,
both in the linear and nonlinear regimes. An outstanding question
concerns the demagnetization processes observed in ferromagnetic
thin films irradiated with femtosecond laser pulses, for which a
clear theoretical explanation is still lacking.

The field of optical control of spins in semiconductor
nanostructures is also a very active research area. It is nowadays
possible to fabricate and optically probe individual semiconductor
quantum dots doped with one or more magnetic impurities
\cite{Besombes}. One of the major interest of this type of
structure is the possibility to control magnetism via optical
processes acting on the charge carriers. Thus, ferromagnetism
becomes optically manipulable on an ultrafast timescale. This is
particulary interesting for the elaboration of future fast-access
magnetic storage devices. We are currently working on quasi one-
and two-dimensional nonparabolic quantum dots containing up to
four electrons and doped with a finite number of localized
magnetic impurities. Within the framework of the CI method and the
Anderson model, we aim at investigating the influence of the
impurities on the energy spectra and oscillator strengths with
special emphasis on the breakdown of the Kohn theorem.

Finally, another procedure that has attracted particular attention
over the last decade is the low-density doping of semiconductor
nanostructures with magnetic impurities such as manganese ions.
The resulting materials (named DMS, for diluted magnetic
semiconductors) can display Curie temperatures as high as 80K
\cite{Wang}, and possibly larger \cite{Dietl}. The spin of the Mn
ions is coupled to the spin degrees of freedom of the electrons
and holes, whose dynamics can be optically excited. DMS thus offer
the possibility of using laser pulses to control the magnetization
dynamics of semiconductor nanostructures.

Given the wealth of fundamental issues and practical applications,
the interplay of charge and spin effects in nanosized objects is
bound to remain a major area of research in the coming years.


%%%%%%%%%%%%%%%%%%%%%%%%%%%%%%%%%%%%%%%%%%%%%%%%%%%%%%%%%%%%%%%%%%%%%%

%\printindex
\end{document}